\documentclass[aps,prx,twocolumn,superscriptaddress,showpacs]{revtex4-1}
\usepackage{amsmath,amssymb,graphics,epsfig,epstopdf,color,verbatim,ulem,braket,tabularx}
\usepackage{multirow}
\usepackage[colorlinks,linkcolor=blue,citecolor=blue,urlcolor=blue,bookmarks=false]{hyperref}
\usepackage{listings}
\usepackage{cancel}
\usepackage{soul}
\usepackage{color}

\begin{document}

\title{Entropy and specific heat of the infinite-dimensional three-orbital Hubbard model}

\author{Changming Yue}
\email{changming.yue@unifr.ch}
\affiliation{Department of Physics, University of Fribourg, 1700 Fribourg, Switzerland}

\author{Philipp Werner}
\email{philipp.werner@unifr.ch}
\affiliation{Department of Physics, University of Fribourg, 1700 Fribourg, Switzerland}

\begin{abstract}
The Hund's coupling in multiorbital Hubbard systems induces spin freezing and associated Hund metal behavior. Using dynamical mean field theory, we explore the effect of local moment formation, spin and charge excitations on the entropy and specific heat of the three-orbital model. In particular, we demonstrate a substantial enhancement of the entropy in the spin-frozen metal phase at low temperatures, and peaks in the specific heat associated with the activation of spin and charge fluctuations at high temperature. We also clarify how these temperature scales depend on the interaction parameters and filling.  
\end{abstract}

\maketitle

\newcommand{\alert}[1]{{\color{red}{#1}}}
 
\section{Introduction}

The three-orbital Hubbard model \cite{annurev_Hund_2013} is relevant for the description of transition metal compounds with partially filled $t_{2g}$ shells \cite{PhysRevLett.92.176403} and for alkali-doped fullerides with half filled molecular orbitals with $t_{1u}$ symmetry \cite{hebard1991,rosseinsky1991,RevModPhys.81.943,PhysRevB.94.155152,PhysRevLett.118.177002}. More generally, it plays an important role in theoretical studies which try to reveal and quantify the correlation effects resulting from the Coulomb interaction in a multi-orbital set-up \cite{prl2008_Werner_spinfreezing,Chan2009,Werner2009,Kita2011,PhysRevLett.107.256401}. For a given electron number, the Hund's coupling $J$ differentiates the energies of atomic configurations with different orbital occupations and spin states. In a lattice environment, it leads to local moment formation and bad metal behavior with a nontrivial filling and temperature dependence \cite{prl2008_Werner_spinfreezing,PhysRevLett.107.256401}. One reason for the dramatic effect of $J$ on the metallic state of multi-orbital systems is the fact that the screening temperature drops exponentially with the magnitude of these local moments \cite{Okada_sd_quench_1973,PhysRevLett.47.737,PhysRevLett.58.843,JPSJ.66.1180,RevModPhys.40.380,Nevidomskyy2009,annurev_Hund_2013}. 

Previous dynamical mean field studies of the three-orbital Hubbard model in the paramagnetic state have demonstrated the existence of a spin-freezing crossover, between a conventional Fermi-liquid type metal in the weakly correlated and strongly doped regime to a bad metal state with frozen magnetic moments near half-filling \cite{prl2008_Werner_spinfreezing,PhysRevLett.107.256401,Hoshino2015}. This crossover regime is characterized by peculiar non-Fermi liquid exponents, such as a self-energy which grows like the square root of frequency in a wide energy window. This crossover has significant effects on the normal state properties of strontium ruthenates \cite{prl2008_Werner_spinfreezing,PhysRevB.58.R10107,PhysRevB.86.195141,PhysRevB.91.195149,PhysRevLett.112.206403,PhysRevLett.116.256401}, iron pnictides \cite{PhysRevB.81.054513,PhysRevLett.108.107201,PhysRevX.3.011006,Haule_2009,Werner_NatPhys2012}, and other correlated materials \cite{PhysRevB.94.245134,lenihan2020entropy}. In models with negative Hund's couplings, relevant for the description of fulleride compounds, an analogous orbital freezing crossover has been observed \cite{PhysRevLett.118.177002,PhysRevB.98.235120}. 

Subsequent work has focused on clarifying the low-temperature properties of these models, and showed that the screening of the orbital and spin moments in the metallic phase eventually leads to Fermi liquid behavior \cite{STADLER2019365,PhysRevLett.115.136401,XYDeng_NatComm_2019}.
 It has also been shown that the enhanced local spin or orbital fluctuations in the crossover regime to the frozen moment state results in unconventional superconducting states at low temperature, while the frozen moment regime itself is susceptible to magnetic or orbital order \cite{Hoshino2015,Steiner2016,Hoshino2016}. This pairing mechanism is interesting because it naturally explains the generic features of the phase diagrams of unconventional superconductors, namely a superconducting dome next to a magnetically ordered phase, and a bad metallic state with non-Fermi liquid properties at elevated temperatures \cite{Hoshino2015}. 

One aspect which has not been systematically studied so far is the fate of the frozen moment regimes and the associated crossovers at elevated temperatures. 
Above some temperature controlled by $J$, we expect the thermal activation of local spin or orbital excitations, which should wash out the freezing effect. 
The entropy of the system should be sensitive to the appearance of long-lived magnetic or orbital moments and can provide new perspectives on freezing-related phenomena. The specific heat, as a closely related quantity, measures fluctuations in the energy, and can thus detect the activation of spin, orbital and charge excitations at high temperatures, as well as the formation of a Fermi liquid state at low temperatures. 

Recent studies have considered the entropy of multi-orbital impurity models in the very low temperature regime, and revealed plateaus associated with the appearance of unscreened moments \cite{2019arXiv190707100H,2019arXiv191013643W}. Here, we present a systematic study of the entropy of the three-orbital lattice system, focusing on the intermediate and high temperature regime and on the case of ferromagnetic Hund's coupling. Our results for the infinitely connected Bethe lattice show relatively broad crossovers associated with the enhancement of the spin entropy 
in the vicinity of Mott phases, below the activation temperature for local spin excitations. Based on an analysis of the atomic problem, we clarify how the activation temperatures for spin and charge excitations and the associated peaks in the specific heat depend on the model parameters.   
 
The paper is organized as follows. In Section~\ref{sec:model} we describe the model and the method used to compute the entropy and specific heat. In Sec.~\ref{sec:jgtr0} we present the results for 
the entropy and specific heat of the three-orbital model, while Sec.~\ref{sec:discussion} contains a discussion and conclusions. Technical details related to the simulation method and data analysis can be found in the Appendices.   
 
\section{Model and Method}
\label{sec:model}

\subsection{Model}

We consider a three-orbital Hubbard model on an infinitely connected Bethe lattice. The local Hamiltonian reads $H_\text{loc}=H_\text{int}-\mu \sum_{\alpha,\sigma} n_{\alpha,\sigma}$ with a density-density interaction term 
\begin{align}
H_{\mathrm{int}}= & \sum_{\alpha}Un_{\alpha,\uparrow}n_{\alpha,\downarrow}+\sum_{\alpha>\beta,\sigma}\Big[ U^{\prime}n_{\alpha,\sigma}n_{\beta,-\sigma}\nonumber\\
 & +\left(U^{\prime}-J\right)n_{\alpha,\sigma}n_{\beta,\sigma}\Big]\label{Hint}.
\end{align}
Here, $\alpha$ labels the orbital, $\sigma$ spin, $n_{\alpha,\sigma}=c^\dagger_{\alpha,\sigma}c_{\alpha,\sigma}$ is the orbital and spin dependent density, $U$ the intra-orbital interaction, $U^{\prime}$ the inter-orbital same-spin interaction, $J$ the Hund's coupling and $\mu$ the chemical potential. We use $U'=U-2J$. Because of the high numerical cost of evaluating the entropy, we do not consider spin-flip and pair-hopping terms. This allows us to use the efficient segment formulation of the hybridization expansion continuous-time Monte Carlo method (CT-HYB) \cite{ctqmc_prl2006,rmp_ctqmc} for the solution of the dynamical mean-field theory (DMFT) \cite{rmp_68_13_dmft_1996} equations. 

In the case of an infinitely connected Bethe lattice, the DMFT solution becomes exact and the self-consistency equations simplify to 
\begin{equation}
\Delta_{\alpha,\sigma}=t^2 G_{\text{imp},\alpha,\sigma},
\label{eq_self}
\end{equation}
where $G_{\text{imp},\alpha,\sigma}$ is the Green's function for the DMFT impurity problem defined by the local term $H_\text{loc}$ and the hybridization functions $\Delta_{\alpha,\sigma}$. The density of states of the noninteracting problem is semi-circular with bandwidth $4t$. In the following, we use $D=2t=1$ as the unit of energy. 

Depending on the filling, temperature, interaction strength and sign of $J$ the three orbital Hubbard model may exhibit antiferromagnetic or ferromagnetic spin or orbital order \cite{Chan2009,Hoshino2015,Hoshino2016,Ishigaki2019}, spin-singlet or spin-triplet superconductivity \cite{Hoshino2015}, as well as symmetry-breaking at the two-particle level \cite{Hoshino2017}. In the present study, we consider $J>0$ and suppress
long range orders, i.e. restrict the solution to the paramagnetic, orbitally degenerate normal state. 

\subsection{Calculation of the entropy}

We compute the entropy per site at density $n$ and temperature $T$ using the formula
\begin{equation}
S(n,T)=S(n,\infty)-\int_{T}^{\infty}\frac{C_{V}(n,T^{\prime})}{T^{\prime}}dT^{\prime},\label{eq:EntropyDef1}
\end{equation}
where the specific heat $C_{V}(n,T)$ is calculated as the derivative of the total energy per site $E_{\text{tot}}(n,T)$ 
with respect to temperature, 
\begin{equation}
C_{V}(n,T)=\frac{\partial E_{\text{tot}}(n,T)}{\partial T}.
\label{Eq_cv}
\end{equation}
We choose the infinite-temperature entropy (rather than the zero temperature value) as the reference, because the CT-HYB method cannot access arbitrarily low temperatures. For a $3$-orbital lattice model with $N$ sites ($N\rightarrow\infty$)
and density $n$, there are $C_{6N}^{nN}$ ways of placing $nN$ electrons on $6N$ spin-orbitals. At $T=\infty$ all these configurations are equally probable and using Stirling's formula we find
\begin{align}
S(n,\infty)&=\lim_{N\rightarrow\infty}\frac{1}{N}\ln C_{6N}^{nN}\nonumber\\
&=
-6\cdot\Big[\frac{n}{6}\ln\frac{n}{6}+\Big(1-\frac{n}{6}\Big)\ln\Big(1-\frac{n}{6}\Big)\Big].
\label{Sinf}
\end{align}
The total energy can be measured in CT-HYB as \cite{prb2007_Haule_qmc}
\begin{equation}
E_{\text{tot}}=\langle H_{\text{int}} \rangle - T\langle k \rangle, 
\end{equation}
where $\langle k \rangle$ is the average total perturbation order of the hybridization expansion and $T$ the temperature ($-T\langle k \rangle$ is the kinetic energy). Since $H_{\text{int}}$ is of density-density type, the interaction energy
can be measured accurately through the sampling of segment overlaps \cite{ctqmc_prl2006}, $\langle H_{\text{int}} \rangle = \sum_{ij} U_{ij} \langle n_i n_j\rangle$, where $i$ and $j$ is a combined spin-orbital index. 
At very high temperatures, $T\gtrsim T_H\equiv U/2$, the energy and specific heat of the system can be well approximated by a Hubbard-I approximation (see Appendix~\ref{app_HubbardI}).  
Hence, in practice, we split the calculation of the entropy into three parts, 
$S(n,T)\approx S(n,\infty)-\int_{T_{H}}^{\infty}\frac{C_{V}^{\text{Hub-I}}(n,T^{\prime})}{T^{\prime}}dT^{\prime}-\int_{T}^{T_{H}}\frac{C_{V}(n,T^{\prime})}{T^{\prime}}dT^{\prime}$,
so that we only have to use CT-HYB simulations in the temperature range
from $T$ to $T_{H}$.

In the noninteracting case, the density of states is temperature independent, so that the total energy ($=$ kinetic energy) can be numerically calculated from the $n$- and $T$-dependent occupation.  

\subsection{Numerical procedure}
We first compute the total energies with high accuracy at many fixed temperatures on a roughly uniform density grid $n \approx 0.1, 0.2, \cdots, 2.9, 3.0$. This density grid is further refined near the integer fillings $n=1, 2, 3$ if these solutions are Mott insulating and hence $\partial E_{\text{tot}}(n,T)/\partial n$ is discontinuous. 
The temperature grid is coarsely spaced for $T\gg J$ and densely spaced for $T\lesssim  J$, with a particularly fine grid around $J/2$ and $J$ in cases where local spin excitations result in a peak in the specific heat near these energy scales (see Appendix~\ref{sec:Atomic-Cv}).

\begin{figure}[t]
\includegraphics[clip,width=3.in,angle=0]{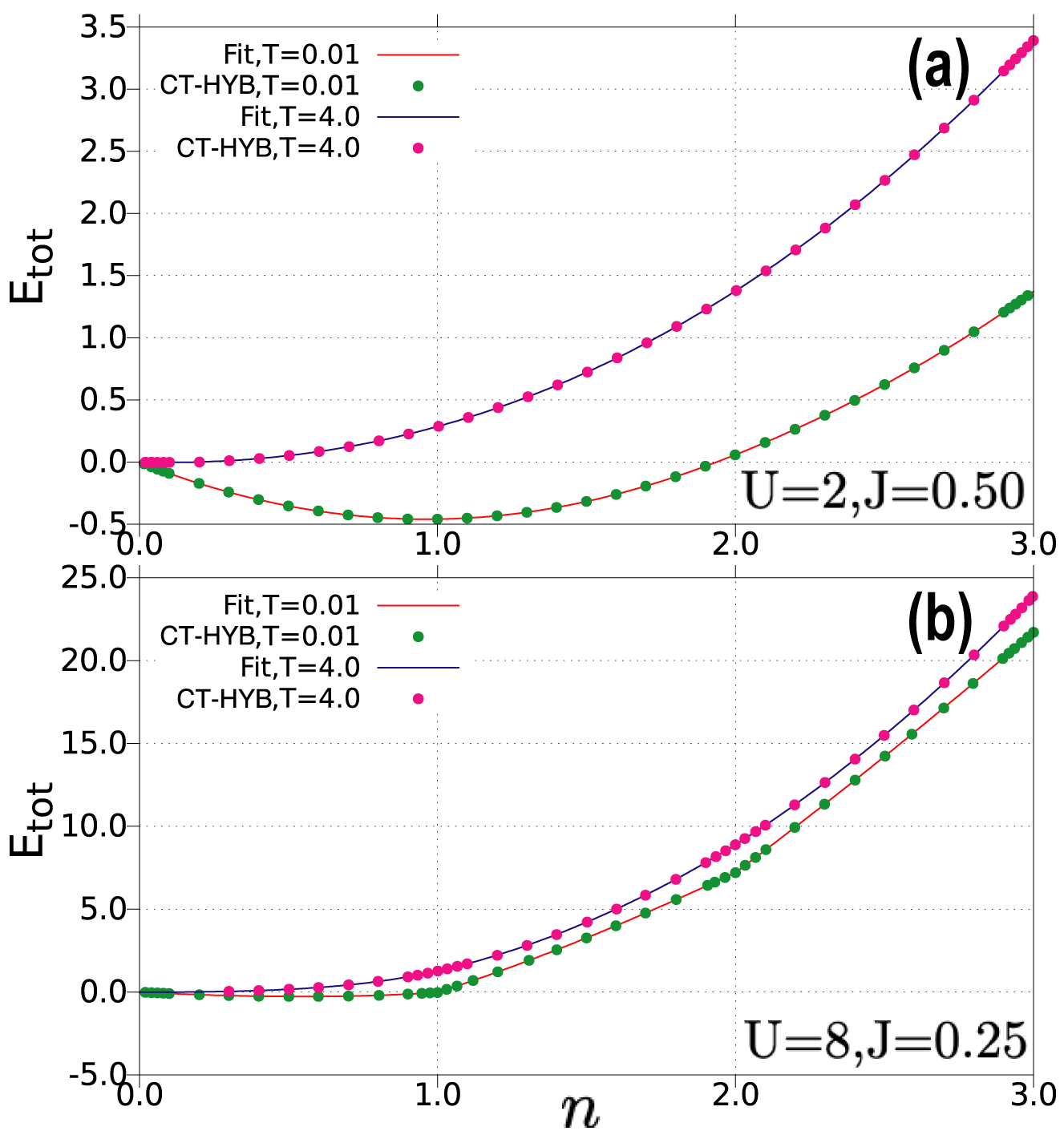}
\caption{(color online). 10-th order polynomial fit along the $n$ axis (lines) to the total energy $E_{\text{tot}}(n,T)$ computed by CT-HYB (dots). Here
we show representative data for low temperature ($T=0.01$) and high temperature ($T=4.0$). Panel (a) shows data for  $U=2.0, J=0.50$ and panel (b) for $U=8.0, J=0.25$. In the latter case, due to the discontinuities at $n=1$ and $n=2$, separate fits are performed for $n\in(0,1)$, $n\in(1,2)$ and $n\in(2,3)$, respectively. 
}
\label{etot_10thPolyFit}
\end{figure}

The $E_{\text{tot}}(n,T)$ data on this non-uniform $n$-$T$ grid are interpolated to a fine and uniform grid using fitting functions. 
More specifically, the data for fixed temperature are interpolated along the $n$ axis using a 10-th order polynomial, $E(n,T_i)=\sum_{m=1}^{10}{c_m(T_i)n^m}$, 
as shown in Fig.~\ref{etot_10thPolyFit}. 
This way, the entropy is obtained on an equidistant grid of 301 $n$-points $\{0.00, 0.01, 0.02, \cdots, 2.99, 3.00\}$.
If Mott insulating solutions exist at $n=1$ and $2$, separate fits are performed for  
$n\in(0,1)$, $n\in(1,2)$ and $n\in(2,3)$, see panel (b). 
In a subsequent step, the resulting $E_{\text{tot}}$ data for fixed filling are fitted along the $T$ axis to eliminate small fluctuations. 
For the low-temperature data points away from Mott phases, we use a polynomial of $T$ \cite{PhysRevB.55.12918} and
calculate $C_V(T)$ by analytically taking the derivatives of this polynomial. If the system is in or very close to a Mott phase,
we fit the data to a function of the form 
\begin{equation}
E_{\text{tot}}(T)=E(0)+\sum_{m=1}^{M} c_n e^{-m\Lambda/T},
\label{sum_exp_fit}
\end{equation}
where $E(0)$, $\Lambda$, and $c_m$ are parameters chosen to minimize the least square errors, and $M$ is chosen as $1/4$ of the number of data points
 \cite{PAIVA2001224,PhysRevLett.82.2342}. At high temperature, the noise is very small and no interpolation is needed in practice, 
although the fitting to Eq.~(\ref{sum_exp_fit}) can still be used.

The simulations are performed using a modified version of the iQIST library \cite{HUANG2015140,iqist}. Since the entropy calculation requires accurate data from very low to very high temperatures, we employ some novel techniques to improve the sampling and measurement efficiency. On the one hand, to reduce the noise in the the standard imaginary-time measurement of the Green's function at high $T$ or large $U$, we use a measurement procedure based on virtual updates, as detailed in Appendix~\ref{sec:virtual_update}. This estimator measures the atomic contribution to the
Green's function when the average expansion order $\langle k \rangle \rightarrow 0$. 
On the other hand, different kinds of global updates are used to avoid trapping in certain configurations. Besides the previously proposed global spin-flip update
\cite{PhysRevB.76.085127,rmp_ctqmc} and the global shift update \cite{state_sampling_prb2019}, we use two additional global updates, which we call double swap update and global shift update. 
The detailed procedures are presented in Appendix~\ref{sec:Global_Updates}. 
The acceptance rates of these global updates can strongly depend on the parameters $n$, $T$, $U$ and $J$. 
In practice, we measure their acceptance rates at the thermalization stage. Those global updates
which are rarely accepted (less than 0.1\%) are disabled during the sampling. 

Finally, the computational effort can be reduced by ensuring a fast convergence of the DMFT loop. For non-integer fillings, instead of using a simple mixing between the Green's functions of subsequent iterations, we use Broyden's method. This procedure has been introduced 
in Ref.~\onlinecite{Broyden_mixing_Rok} and it can lead to a speed-up by up to a factor of three in well-behaved cases. 

\section{Results}
\label{sec:jgtr0}

\subsection{Entropy surfaces}

We consider two representative parameter sets for the interacting three-orbital system: (i) $U=2$, $J=0.5$ and (ii) $U=8$, $J=0.25$. The first choice corresponds to a model which at zero temperature is Mott insulating at half-filling but metallic away from half-filling, with a $T$-dependent crossover from a spin-frozen to a Fermi liquid metal phase \cite{prl2008_Werner_spinfreezing,PhysRevLett.107.256401}. The second choice corresponds to a strongly correlated system that is Mott insulating at fillings $n=1$, $2$, and $3$. The smaller $J/U$ ratio in this model results in a clear separation between the temperature scales associated with the activation of spin and charge degrees of freedom. In order to reveal the correlation effects, we will also compare the results from models (i) and (ii) to the noninteracting system.  

\begin{figure*}[htp]
\includegraphics[clip,width=0.835\paperwidth,angle=0]{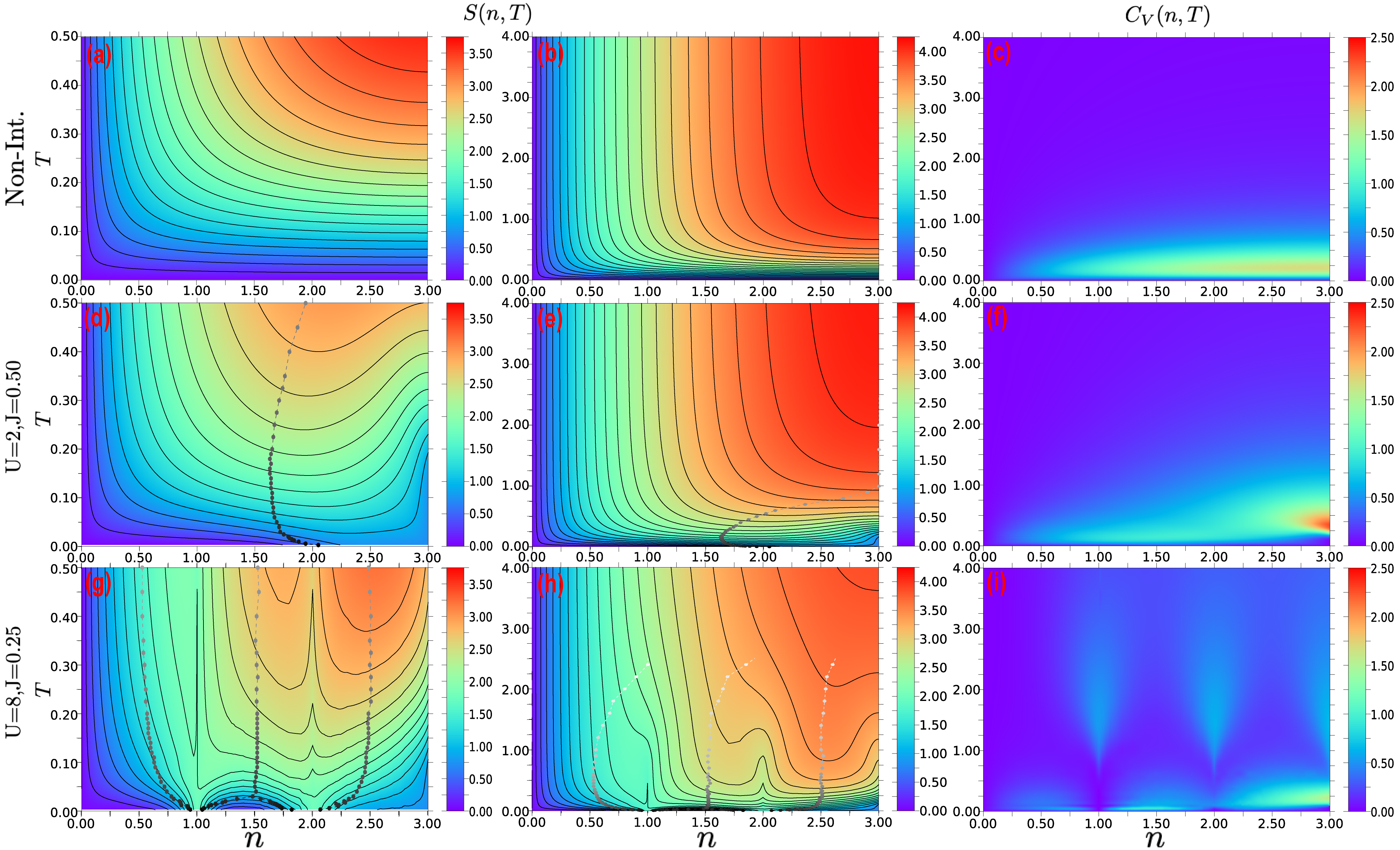}
\caption{ (color online). Contour maps of the entropy per site $S(n,T)$ and color map of the specific heat per site $C_V(n,T)$ as a function of filling and temperature. 
Panels (a-c) are for the non-interacting system, panels (d-f) for model (i) ($U=2$, $J=0.50$) and panels (g-i) for model (ii) ($U=8$, $J=0.25$), respectively.
Dashed lines with dots 
in (d-e) and (g-h) show the locations of the maxima of $\Delta\chi_{\text{loc}}$ (determined as a function of $n$ at fixed $T$). The gray scale of these points represents $-\log_{10}\Delta\chi_{\text{loc}}$.
}
\label{schematic}
\end{figure*}

Contour maps of $S(n,T)$ are shown in the left two columns of Fig.~\ref{schematic}, with the first row corresponding to the noninteracting model ($U=J=0$), the second row to model (i) and the third row to model (ii). For very low fillings, interaction effects are negligible, and the entropy contours of the three models are almost identical. Clear differences however appear near half-filling. Here, at low temperature, the interacting models are in a Mott insulating state with three electrons per site. While the Hund's coupling results in aligned spins, the orientation of the resulting spin-3/2 moments is random in our paramagnetic simulations, so that the system exhibits a $\ln 2$ entropy per site (note that the spin rotation invariance is broken in the model with density-density interactions).  An even larger enhancement of the low-$T$ entropy is found in model (ii) near filling $n=2$ and $n=1$. Here, due to the orbital degree of freedom, there are six degenerate states with spin-1 and spin-1/2, respectively, resulting in a $\ln 6$ entropy per site at low temperatures. 

\begin{figure}[htp]
\includegraphics[clip,width=3.4in,angle=0]{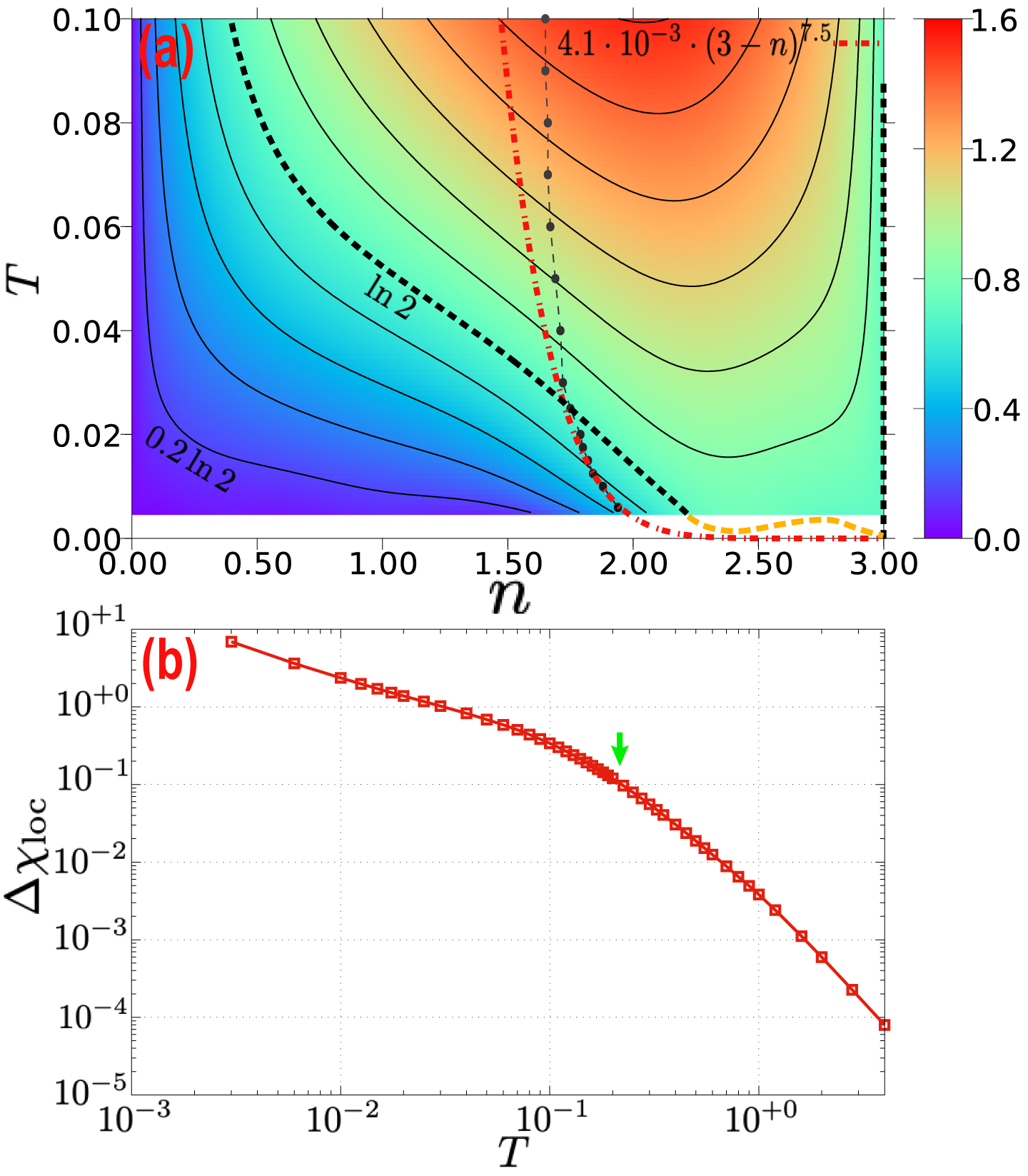}
\caption{ (color online). (a) Contour map of the entropy of model (i) ($U=2, J=0.50$) at low $T$.
The increment of the contour lines is $\frac{\ln 2}{5}$. The thick black dashed lines (including the one at $n=3$) show the $\ln 2$ contour, 
while the orange dashed line is a guide to the eye.
The points with variable gray scale indicate the locations of the maxima of $\Delta \chi_{\text{loc}}$ at fixed $T$ (spin-freezing crossover line).
The red dash-dotted line is a fit of the form $a\cdot(3-n)^\alpha$ to this line, using the 4 data points with the lowest $T$.
(b) The magnitude of $\Delta\chi_{\text{loc}}$ along the spin-freezing crossover line, plotted as a function of temperature. The green arrow indicates the temperature corresponding to the spin excitation peak in $C_V(n,T)$.
}
\label{S_lowT}
\end{figure}

A zoom into the low-temperature behavior of model (i) is shown in the top panel of Fig.~\ref{S_lowT}, with the entropy contour line corresponding to $\ln 2$ highlighted (thick dashed line). Evidently, the entropy of the correlated metal is enhanced at low temperatures ($T\lesssim 0.2$) and low dopings ($2\lesssim n \lesssim 3$) compared to a conventional Fermi liquid, and comparable in magnitude to the entropy of the $n=3$ Mott insulator. In fact, in a Fermi liquid the entropy behaves as $S(n,T)=\gamma(n)T$, with $\gamma(n)=\lim_{T\rightarrow 0}C_V(n,T)/T$, and in the absence of spin-freezing we would expect $\gamma(n)\propto 1/|n-3|$. This scaling has been explicitly demonstrated for the single-band Hubbard model in Ref.~\onlinecite{Werner2007doped}. It implies entropy contours which are straight lines emanating from the point $(n=3,T=0)$. For large enough doping ($n\lesssim 1.5$) the low-entropy contours in Fig.~\ref{S_lowT}(a) indeed exhibit this expected Fermi-liquid behavior, but for $n\gtrsim 1.5$ one observes a downturn of the entropy lines and the formation of a high-entropy plateau with a value of approximately $\ln 2$. This is the manifestation of spin-freezing in the doping and temperature dependence of the entropy. Indeed, this filling and temperature range corresponds to the spin-frozen regime of this three-orbital system, which has been identified in previous DMFT investigations via an analysis of the self-energy \cite{prl2008_Werner_spinfreezing}, local spin susceptibility \cite{Hoshino2015} and quasi-particle weight \cite{PhysRevLett.107.256401}. 

At very low temperatures (not accessible with CT-HYB) the frozen moments will be screened \cite{Nevidomskyy2009} and the entropy of the resulting strongly renormalized Fermi liquid will go to zero as $T\rightarrow 0$. 
This physics has been demonstrated and explored in recent DMFT studies employing NRG impurity solvers \cite{PhysRevLett.115.136401}, and in related works on multi-orbital impurity models \cite{2019arXiv190707100H,2019arXiv191013643W}.

\begin{figure*}[htp]
\includegraphics[clip,width=0.8\paperwidth,angle=0]{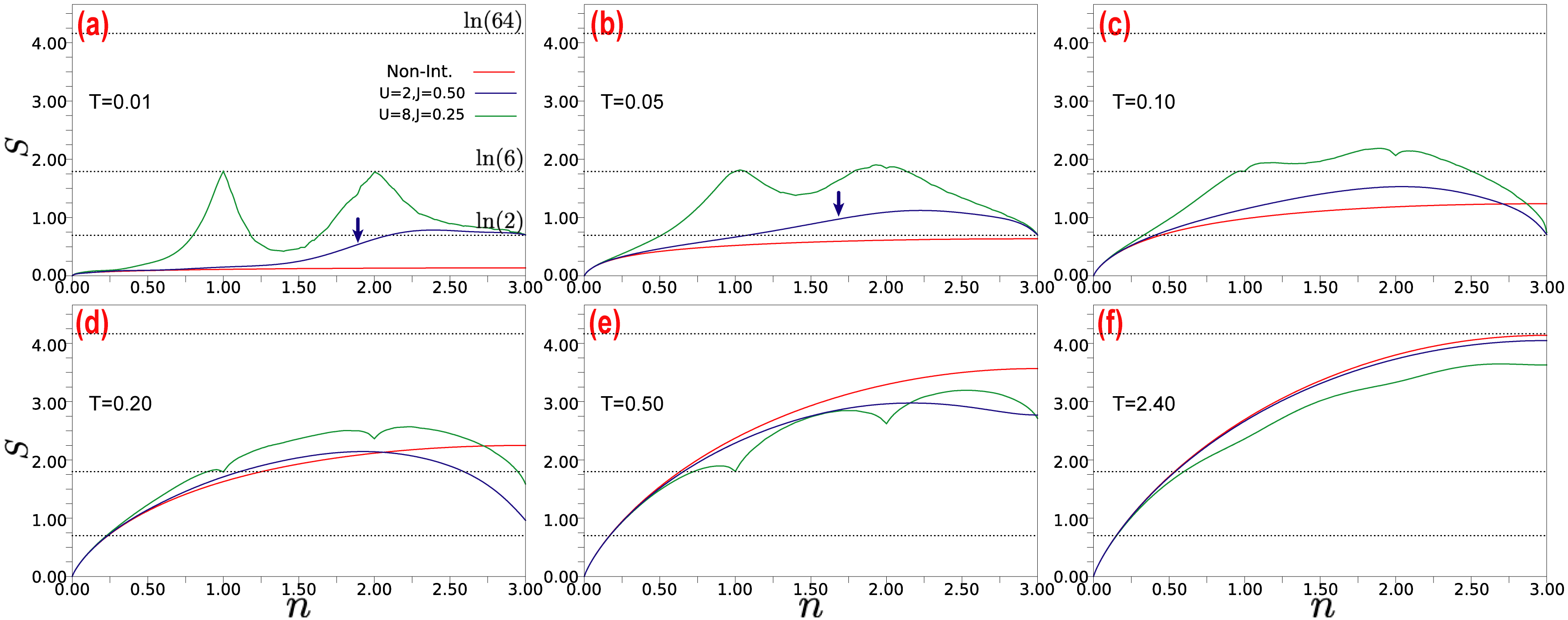}
\caption{ (color online). Filling dependence of the entropy per site $S(n,T)$ at six different temperatures, $T=0.01, 0.05, 0.10, 0.20, 0.50, 2.40$, respectively. The red lines are for the non-interacting system, the  
blue lines for model (i) $(U=2, J=0.50$), and the green lines for model (ii) ($U=8, J=0.25$). 
The blue solid arrows in panels (a) and (b) indicate the filling at which $\Delta\chi_\text{loc}$ reaches its maximum (see Fig.~\ref{ChilocFixT_n}(a,b)).
}
\label{EntropyFixT_n}
\end{figure*}

Cuts of the entropy surfaces at fixed temperatures are shown in Fig.~\ref{EntropyFixT_n}. The blue line in panel (a) illustrates the spin-freezing related increase of the entropy in model (i) and the approximate $\ln 2$ plateau in the filling range $2\lesssim n \lesssim 3$. The low temperature results for both models furthermore confirm the theoretically expected entropy values of the Mott phases at integer fillings (see horizontal dashed lines). By comparing the entropies for different temperatures to the noninteracting result, we see how local moment formation in and near the Mott phases leads to a substantial increase in the entropy at low temperatures, while the suppressed charge fluctuations in the interacting systems reduce the entropy at very high temperatures. 

\begin{figure*}[htp]
\includegraphics[clip,width=0.8\paperwidth,angle=0]{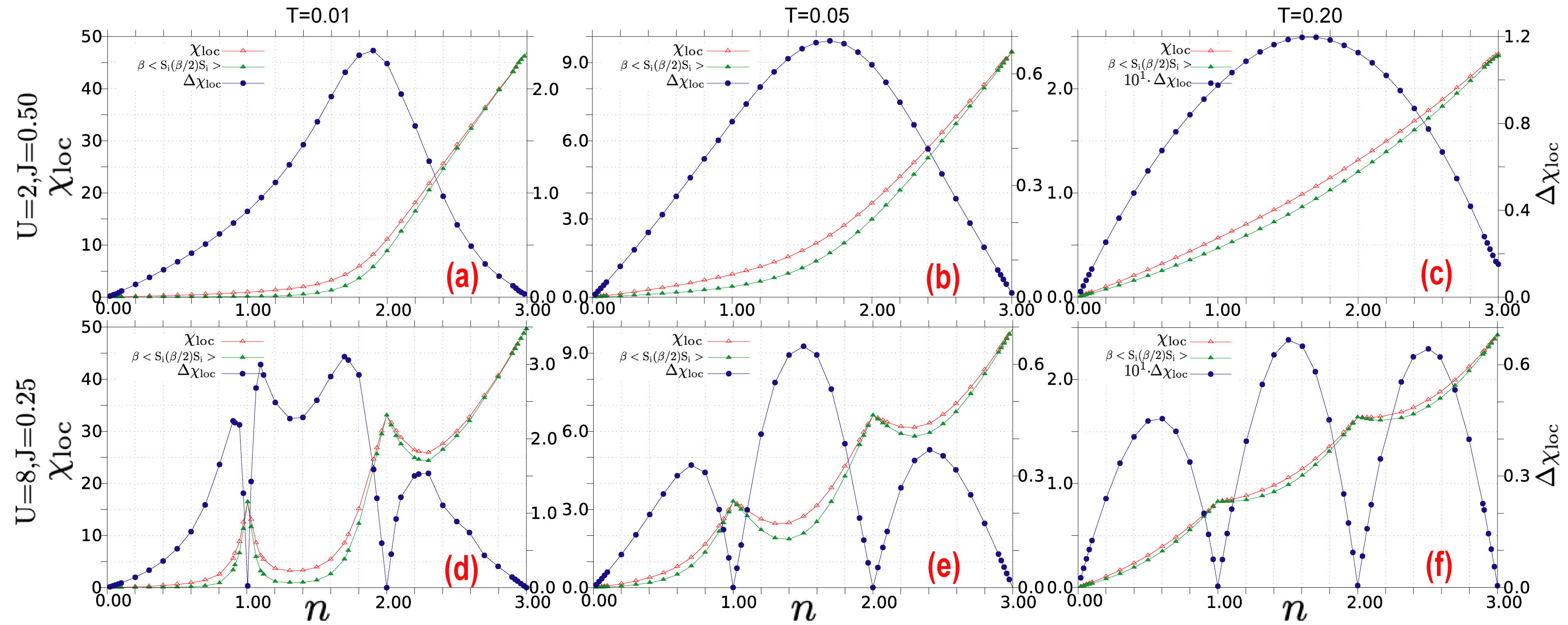}
\caption{ (color online). Filling dependence of the local spin susceptibility at $T=0.01$ (first column), $T=0.05$ (second column), and $T=0.20$ (third column). 
The first row is for model (i) ($U=2$, $J=0.50$) and the second row for model (ii) ($U=8$, $J=0.25$). 
The red line shows $\chi_\text{loc}$, the green line the frozen moment contribution $\beta\langle S_i(\beta/2)S_i(0)\rangle$, and the blue line the fluctuating contribution $\Delta\chi_\text{loc}$.
}

\label{ChilocFixT_n}
\end{figure*}

\begin{figure*}[htp]
\includegraphics[clip,width=0.8\paperwidth,angle=0]{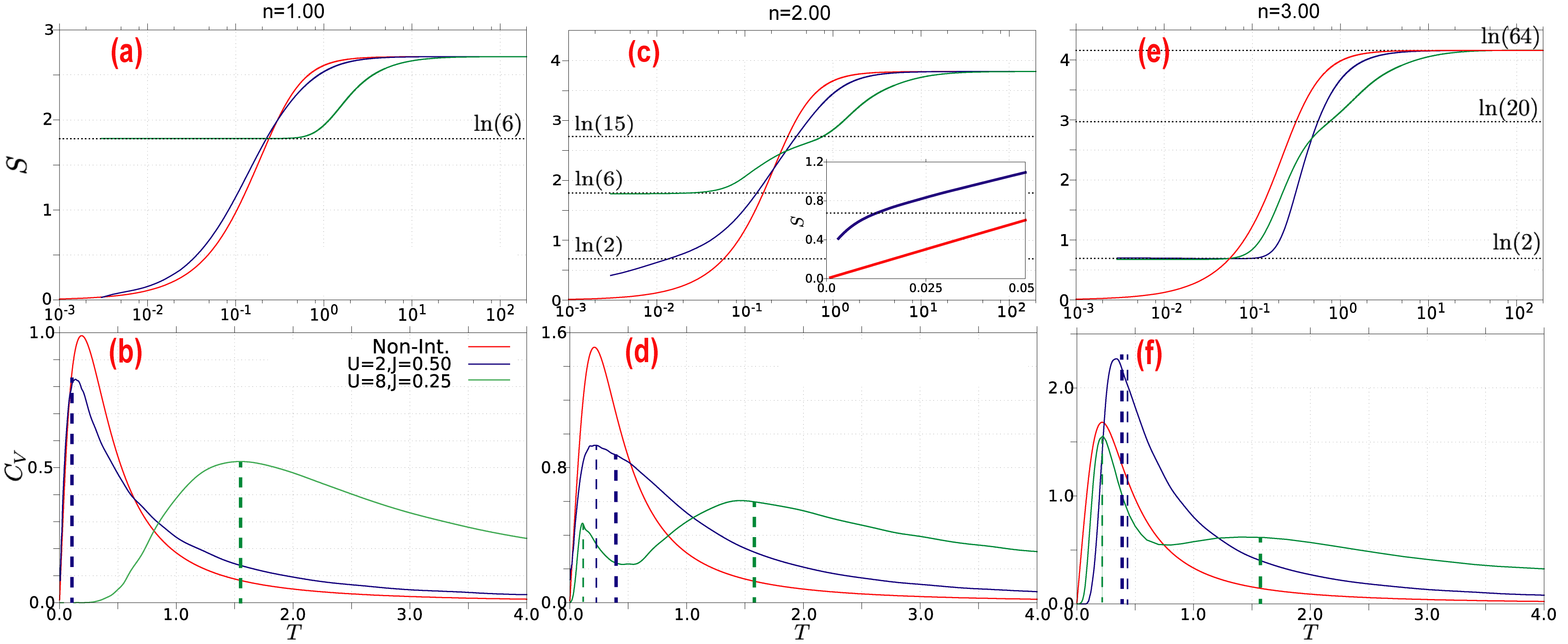}
\caption{ (color online). Entropy (first row) and specific heat (second row) as a function of $T$ for model (i) and model (ii) at the integer fillings $n=1$ (first column), $n=2$ (second column) and  
$n=3$ (third column).
Inset of panel (c): Entropy at $n=2$ plotted on a linear $T$ instead of a $\log_{10}T$ axis to show the onset of the Fermi liquid crossover in model (i).
In panels (d) and (f), the thin dashed lines locate the positions of $T_{\text{Hund}}$, while the thick dashed lines locate the positions of $T_{\text{charge}}$, as listed in Table~\ref{tab_peaks}.
In panel (b), the thin dashed line shows the more accurate estimate of the charge peak position at $n=1$,  $T_\text{charge}^{(n=1)}=0.214(U-3J)$ (see Appendix \ref{sec:Cv_n_eq_1_Ueff}). 
}
\label{EntropyFixn_T}
\end{figure*}

\subsection{Spin freezing}

As shown in Fig.~\ref{ChilocFixT_n} (top row), at low temperatures, the local spin susceptibility 
\begin{equation}
\chi_\text{loc}=\int_0^\beta d\tau \langle S_z(\tau) S_z(0)\rangle
\end{equation} 
for model (i) starts to increase rapidly with filling around $n\approx 1.8$. This increase of $\chi_\text{loc}$ is the direct manifestation of spin freezing. In Ref.~\onlinecite{Hoshino2015} the location of the spin freezing crossover has been defined by measuring the maximum in the fluctuating contribution to the local spin susceptibility, 
\begin{equation}
\Delta\chi_\text{loc}=\chi_\text{loc}-\beta\langle S_z(\beta/2)S_z(0)\rangle. 
\end{equation}
This quantity is shown by the blue lines in Fig.~\ref{ChilocFixT_n} and exhibits a peak at low temperatures in the filling region where the local spin susceptibility starts to grow, i.e. in the region where local moments start to form. 

We indicate the thus defined spin-freezing crossover points by the grey dots in Fig.~\ref{schematic} and Fig.~\ref{S_lowT}. 
In Fig.~\ref{EntropyFixT_n}(a,b) we show the filling corresponding to the peak in $\Delta\chi_\text{loc}$ for model (i) by the blue arrow.  It is clear from 
these dotted lines and arrows
that the spin-freezing crossover indeed explains the enhancement of the entropy near $n\approx 2$ at low temperature. 

Note that the amplitude and sharpness of the peak in $\Delta\chi_\text{loc}$ decreases with increasing temperature, such that a reasonably sharp crossover point can only be defined for $T\lesssim 0.2$. 
In Fig.~\ref{S_lowT}(b) we plot $\Delta\chi_\text{loc}$ as a function of $T$ along the spin-freezing line. Around $T\approx 0.2$ there is a kink in this log-log plot, which marks a temperature $T_\text{sf}$ that separates the temperature range with a sharp spin-freezing crossover ($T\lesssim T_\text{sf}$) from the higher temperature region with no well-defined spin-freezing crossover. As we will see below, this temperature corresponds to the activation temperature for local spin excitations in model (i) at $n=2$ (green arrow in Fig.~\ref{S_lowT}(b)). 
A clear effect of the spin-freezing crossover on the entropy contours is however only visible for ten times lower temperatures, since the spin-freezing line crosses the $\ln 2$ contour near $T\approx 0.025$.  

As a side remark, we note that the $T$-$n$ behavior of our spin-freezing line seems to be different from the results reported in Fig.~15 of Ref.~\onlinecite{STADLER2019365}, which shows a spin-screening temperature of the form $(3-n)^\alpha$ with an exponent $\alpha$ in the range from $2$ to $3$ (depending on parameters). A power-law fit to the lowest-temperature points of our spin-freezing crossover line yields an exponent $\alpha \approx 7.5$-8.5. We see two possible reasons for this discrepancy: (a) the actual power-law scaling of this crossover line may be restricted to temperatures which are lower than the $T=0.005$ reached in our study, and (b) the exponent $\alpha$ is likely larger in the model with density-density interactions considered here, than in the spin-rotation invariant model of Ref.~\onlinecite{STADLER2019365}. In fact, it was previously shown that the model with density-density interactions has a more extended spin-freezing region and a sharper onset of the spin-frozen regime \cite{Hoshino2015}. 

In model (ii), with three Mott insulating solutions at low temperatures, local moments form near $n=1$, $2$ and $3$, resulting in several maxima in $\Delta\chi_\text{loc}$ (see bottom row of Fig.~\ref{ChilocFixT_n}). As temperature increases, these maxima become weaker and shift away from the integer fillings, which eventually results in three weak humps near fillings 0.5, 1.5 and 2.5. The spin-freezing crossover associated with the $n=3$ Mott state, evident in the doping evolution of $\chi_\text{loc}$, also exists around $n\approx 2$, similar to the case of model (i) (compare upper and lower panels in Fig.~\ref{ChilocFixT_n}), but it is masked in $\Delta\chi_\text{loc}$ by the presence of the $n=1$ and $n=2$ Mott states. Hence, the meaning of the gray dots in model (i) and (ii) is different, especially at the higher temperatures, and only in model (i), where the spin freezing occurs in a filling and interaction regime that is clearly separated from the Mott solutions, should we talk about Hund metal behavior. 

Concerning local moment formation, it is interesting to note that in the strongly correlated regime of model (ii), the fidelity susceptibility allows an even more sensitive detection of such moments than $\Delta\chi_\text{loc}$ \cite{Huang2016}.   

\subsection{Entropy and specific heat at integer fillings}

The data in Fig.~\ref{EntropyFixT_n} show that the entropy of the Mott insulating solutions at $n=1,2$ and $3$ is pinned at $\ln 6$ and $\ln 2$, respectively,  below a temperature which depends on $J$, $U$ and filling. In the insulating case, we expect that local spin excitations become relevant at a temperature scale determined by $J$, while charge excitations contribute to the entropy above a temperature scale determined by $U$ and $J$. In the top panels of Fig.~\ref{EntropyFixn_T} we plot the $T$-dependence of the entropy 
for the integer fillings $n=1$, $2$ and $3$. In the $n=1$ case, the Hund's coupling should play a minor role and indeed we observe only a single crossover in the entropy from the low-temperature value of $\ln 2$ (0) for the Mott insulator (metal) to the infinite-temperature value of $\ln 6 + 5\ln \frac{6}{5}$ (Eq.~(\ref{Sinf})). 
This crossover is associated with charge excitations, and thus occurs at a higher temperature in the Mott insulating case (green line, model (ii)).

For $n=2$, the entropy curve for parameter set (ii), with a clear separation between $J$ and $U$, exhibits an intermediate temperature plateau that likely corresponds to a regime with activated spin, but still frozen charge fluctuations. To support this interpretation, we plot in panel (c) as a horizontal line the $\ln 15$ value, which corresponds to the entropy of the 15 atomic states with two electrons in three orbitals. 
A similar but less prominent intermediate plateau is also evident for $n=3$, see panel (e). At this filling, there are 20 local states, so that the intermediate plateau appears around $\ln 20$. 

Model (i), which is also Mott insulating for $n=3$, exhibits what appears to be a single crossover from the $\ln 2$ entropy of the Mott state to the $\ln 64$ entropy of the infinite temperature state. As we will see below, this is because spin and charge excitations are activated in the same temperature range for this parameter set. More interesting is the behavior of the entropy of model (i) at $n=2$, which is in the metallic phase. As panel (c) shows, the entropy per site of this metal remains above $\ln 2$ for temperatures down to about $T\approx 0.01$, and then rapidly drops to zero. This rapid drop is masked by the logarithmic scale in the main panel, but is clearly evident in the inset, which uses a linear temperature axis. The steep increase of the entropy at low temperatures (compared to the noninteracting model, red line) is the result of spin-freezing, i.e. of the emergence of long-lived composite spin-1 moments in the metal phase, as discussed in the previous section. 
Conversely, the rapid drop of $S(n=2,T)$ below $T\approx 0.01$ is due to the screening of these local moments, and the formation of a low-temperature Fermi liquid states. At the lowest temperatures accessible to our CT-HYB simulations, it is possible to see the onset of this screening. With the help of NRG solvers, the complete crossover to the Fermi liquid ground state has recently been demonstrated \cite{PhysRevLett.115.136401,STADLER2019365}.

\begin{figure*}[htp]
\includegraphics[clip,width=0.8\paperwidth,angle=0]{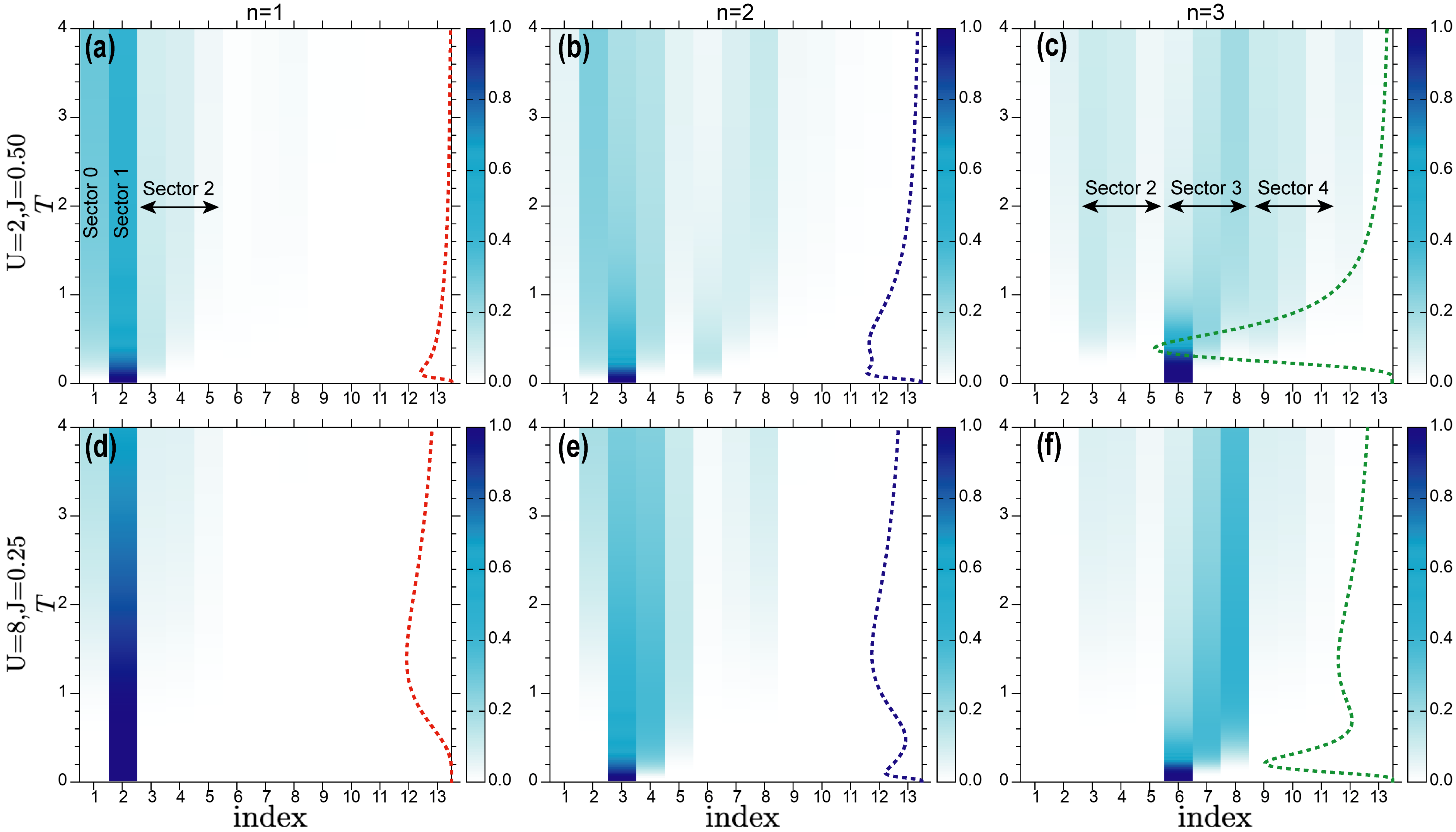}
\caption{ (color online). Temperature dependent probability distribution (Eq.~\eqref{eq:prob_atomic}) of the atomic multiplets (see Table~\ref{tab_eigenstates}) and corresponding atomic specific heat $C_V(T)$ calculated
by the numerical derivative of the atomic energy (Eq.~\eqref{eq:E_int_atomic}). The first row is for model (i) ($U=2,J=0.50$) and the second row for model (ii) ($U=8,J=0.25$). 
The first, second, and third columns are for filling $n=1$, $n=2$ and $n=3$, respectively. 
}
\label{atomic_prob_and_Cv}
\end{figure*}

It is clear from the relation between the entropy and specific heat (Eq.~\eqref{eq:EntropyDef1}) that the crossovers in the temperature dependence of $S(n,T)$ are associated with peaks in $C_V(n,T)$, as is explicitly shown in the lower panels of Fig.~\ref{EntropyFixn_T}. Since the specific heat measures energy fluctuations, the latter peaks reflect the different types of excitations in the system. 
$C_V(T)$ has been previously computed and discussed for the half-filled one-band Hubbard model. A two-peak structure 
\cite{PhysRevB.12.2260,PhysRevB.55.12918,rmp_68_13_dmft_1996,PAIVA2001224}
has been observed in the strongly correlated metallic phase at $U=2D$. The narrower low-energy peak is associated with the emergence (respectively screening) of a local moment, and thus similar to the peak associated with the crossover from the Fermi liquid to the spin frozen state discussed above.
The characteristic energy scale for this peak is the renormalized Fermi energy $\epsilon_F^*=ZD$, 
with $Z$ the quasi-particle weight. 
The broader peak at high energies $T\sim U$ has been attributed to charge fluctuations.

In the three orbital case, the situation is more complex. First of all, $\epsilon_F^*$ is substantially reduced near half-filling, because the formation of the Fermi liquid state requires the screening of a large composite moment  (spin $\sim 3/2$)\cite{Nevidomskyy2009}. Furthermore, there are spin and orbital moments. In the paramagnetic phase of the 1/3 filled three-band Hubbard model
with rotationally invariant interaction, four different low-energy scales have been identified \cite{XYDeng_NatComm_2019}, which mark the onset and the completion
of screening in the orbital and spin channels, respectively. 

While in a Mott state, we expect a charge excitation peak at an energy scale determined by $U$, similar to the one-band case, there should also be a lower energy feature associated with local spin excitations. We will call such a peak in the specific heat the ``Hund excitation" peak. 

\begin{figure*}[htp]
\includegraphics[clip,width=0.8\paperwidth,angle=0]{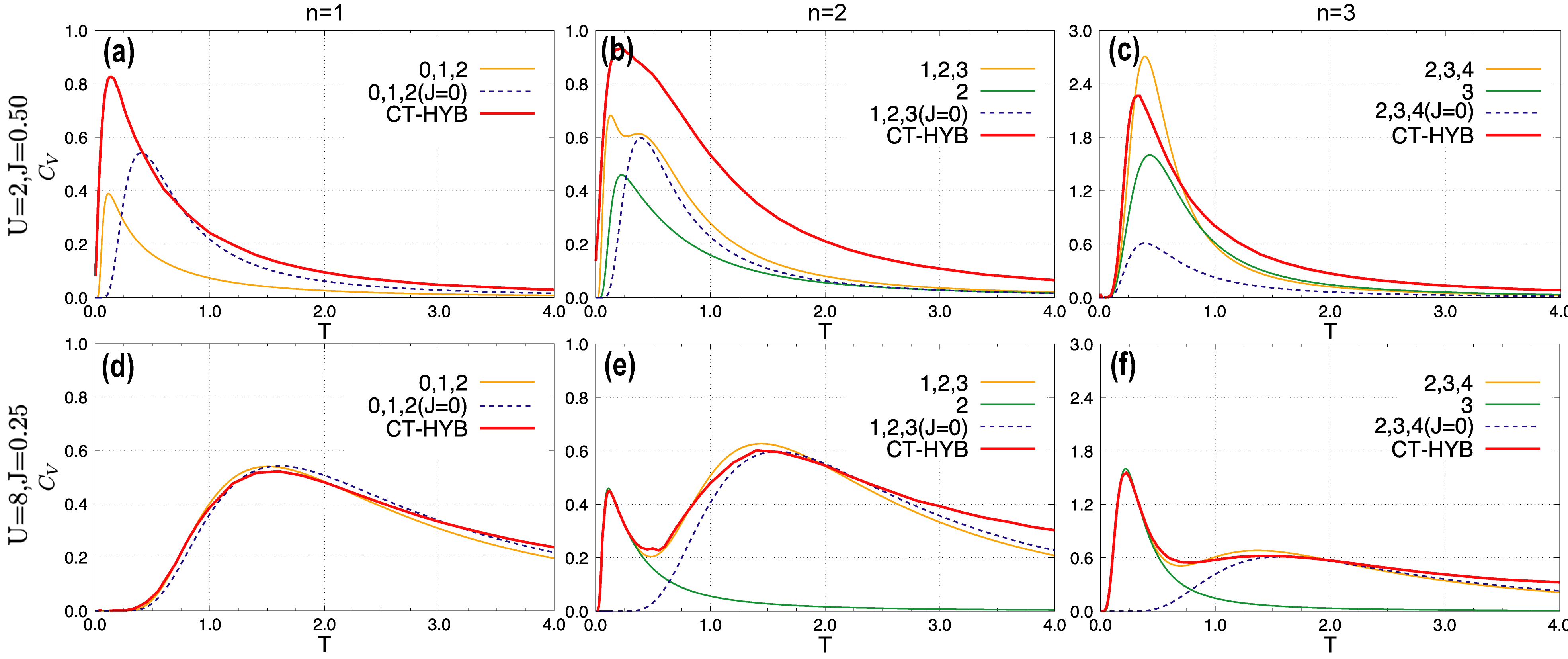}
\caption{ (color online). 
Comparison between the specific heat of the DMFT solution obtained by CT-HYB and the atomic specific heat calculated by applying sector truncations (see Appendix \ref{sec:Atomic-Cv}).
The first row is for model (i) ($U=2,J=0.50$) and the second row for model (ii) ($U=8,J=0.25$).
The first, second and third column are for the integer fillings $n=1$, $n=2$ and $n=3$, respectively.
Numbers in the legends refer to the corresponding sectors listed in Table~\ref{tab:Atomic-Eigenstates}. The orange lines represent calculations which take into account both $U$ and $J$, while the blue dashed lines show the charge peak obtained by setting $J=0$. The green lines at $n=2$ and $n=3$ show the Hund peak calculated by considering only the dominant sector.
}
\label{cv_atomic_full_and_subspace_ctqmc_n1n2n3}
\end{figure*}

\subsection{Hund and charge excitations}

In the following, we will use an analysis of the specific heat in the atomic limit to identify the characteristic energy scales for the Hund and charge excitations at filling $n=1$, $2$, and $3$.  

The atomic Hamiltonian $H_\text{loc}=H_\text{int}-\mu \sum_{\alpha,\sigma} n_{\alpha,\sigma}$ (see Eq.~\eqref{Hint}) 
can be solved by exact diagonalization. The eigenstates may be categorized into sectors, or subspaces, labelled by the  occupation number $N_{\Gamma}$. All eigenvectors, eigenvalues, and their degeneracies are listed in Table~\ref{tab:Atomic-Eigenstates}. The chemical 
potential $\mu$ needs to be properly adjusted to ensure the correct average filling $n$. 
\begin{table*}
\caption{\label{tab:Atomic-Eigenstates}
Eigenvectors and eigenvalues of $H_{\mathrm{int}} -\mu\sum_{\alpha,\sigma} n_{\alpha,\sigma}$. 
We classify the sectors according to their occupation $N_{\Gamma}$. $a\ne b\ne c = 1,2,3$ is the orbital index,
while $\sigma=\uparrow,\downarrow$ denotes the spin.
}
\begin{ruledtabular}
\begin{tabular}{cccccc}
Sector and $N_{\Gamma}$ & Degeneracy & Index & Label & Eigenvector $\Gamma$ & Eigenvalue\tabularnewline
\hline 
0 & 1 & 1 & $\phi_{0}$ & $|\rangle$ & 0\tabularnewline
\hline 
1 & 6 & 2 & $\phi_{1}$ & $|a\sigma\rangle$ & $-\mu$\tabularnewline
\hline 
\multirow{3}{*}{2} & 6 & 3 & $\phi_{2G}$ & $|a\sigma,b\sigma\rangle$ & $U-3J-2\mu$\tabularnewline
 & 6 & 4 & $\phi_{21}$ & $|a\sigma,b\overline{\sigma}\rangle$ & $U-2J-2\mu$\tabularnewline
 & 3 & 5 & $\phi_{22}$ & $|a\uparrow a\downarrow\rangle$ & $U-2\mu$\tabularnewline
\hline 
\multirow{3}{*}{3} & 2 & 6 & $\phi_{3G}$ & $|a\sigma b\sigma c\sigma\rangle$ & $3U-9J-3\mu$\tabularnewline
 & 6 & 7 & $\phi_{31}$ & $|a\sigma b\sigma c\overline{\sigma}\rangle$ & $3U-7J-3\mu$\tabularnewline
 & 12 & 8 & $\phi_{32}$ & $|a\uparrow a\downarrow,b\sigma\rangle$ & $3U-5J-3\mu$\tabularnewline
\hline 
\multirow{3}{*}{4} & 6 & 9 & $\phi_{4G}$ & $|a\uparrow a\downarrow,b\sigma c\sigma\rangle$ & $6U-13J-4\mu$\tabularnewline
 & 6 & 10 & $\phi_{41}$ & $|a\uparrow a\downarrow,b\sigma c\overline{\sigma}\rangle$ & $6U-12J-4\mu$\tabularnewline
 & 3 & 11 & $\phi_{42}$ & $|a\uparrow a\downarrow,b\uparrow b\downarrow\rangle$ & 6$U-10J-4\mu$\tabularnewline
\hline 
5 & 6 & 12 & $\phi_{5}$ & $|a\uparrow a\downarrow,b\uparrow b\downarrow,c\sigma\rangle$ & $10U-20J-5\mu$\tabularnewline
\hline 
6 & 1 & 13 & $\phi_{6}$ & $|1\uparrow1\downarrow,2\uparrow2\downarrow,3\uparrow3\downarrow\rangle$ & $15U-30J-6\mu$\tabularnewline
\end{tabular}
\end{ruledtabular}
\label{tab_eigenstates}
\end{table*}

The partition function reads 
\begin{align}
& Z =\mathrm{Tr}e^{-\left(H_{\mathrm{int}}-\mu\sum_{\alpha,\sigma}n_{\alpha,\sigma}\right)/T}=\sum_{s=0}^{6}\sum_{i\in s}\ensuremath{d_{s,i}}e^{-E_{s,i}/T},
\label{Z_full_atomic}
\end{align}
where the sum is taken over all eigenstates labelled by the sector number $s$ and an index $i$ referring to a subgroup of states within this sector ($d_{s,i}$ denotes the corresponding degeneracies, see Table~\ref{tab:Atomic-Eigenstates}). 
The probabilities of the atomic multiplets are
\begin{equation}
p_{s,i}(T)=\frac{d_{s, i} e^{-E_{s, i} / T}}{Z}.
\label{eq:prob_atomic}
\end{equation}
To avoid numerical problems due to large exponentials, we employ an energy shift \cite{PhysRevB.86.075153}
$E_{s,i}\rightarrow E_{s,i}-E_0^{\mu}$ where $E^0_{\mu}$ denotes the lowest energy for all eigenstates at a certain $\mu$.

The total energy  
\begin{equation}
E_{\text{tot}}(n,T)=\frac{1}{Z}\sum_{s=0}^{6}\sum_{i\in s}\ensuremath{d_{s,i}}E_{s,i}e^{-E_{s,i}/T}+\mu n
\label{eq:E_int_atomic}
\end{equation} 
yields the electronic specific heat per atom via Eq.~\eqref{Eq_cv},
which we numerically evaluate by a finite difference method on a fine enough $T$ grid. 

In Fig.~\ref{atomic_prob_and_Cv} we plot $p_{s,i}(T)$ and $C_V(T)$ for the three integer fillings 
$n=1$, $2$ and $3$, respectively. The top panels are for model (i) and the bottom panels for model (ii). 
The $x$-axis is the index of the states defined in the $3$rd column of Table~\ref{tab:Atomic-Eigenstates}, 
while the temperature-dependent probability of these states is indicated by the color. 
The figure shows that the peaks in $C_V(T)$ are directly related to the temperature evolution of the atomic probabilities. 

Let us first discuss the results for model (ii), which has a large $U$ and $J\ll U$. In the case of $n=1$ (panel (d)), the specific heat exhibits a single broad peak near $T\approx 1.5$. This peak correlates with the population of states in the $N_\Gamma=0$ and $N_\Gamma=2$ sectors and thus originates from thermally activated charge excitations. Local spin excitations do not play a role at low temperatures, since the dominant $N_\Gamma=1$ sector does not admit such excitations. As shown in Appendix~\ref{sec:charge_peak_ntot_1}, if we neglect $J$ and only consider the sectors $0$, $1$ and $2$, the charge excitation peak in $C_V(T)$ is located at $T_\text{charge}^{(n=1)}=0.201U$, which agrees well with the full calculation for model (ii). 

At filling $n=2$ (panel (e)), the specific heat shows two peaks, a low-energy peak near $T=0.1$ and a high-energy peak near $T=1.5$. As the distribution of the probabilities $p_{s,i}(T)$ clearly reveals, the low-energy peak originates from excitations within the dominant $N_\Gamma=2$ sector, and is thus associated with Hund excitations. The higher energy peak, on the other hand, correlates with the population of the neighboring $N_\Gamma=1$ and $N_\Gamma=3$ sectors, and thus is a charge excitation peak analogous to the one discussed for $n=1$. The calculation in Appendix ~\ref{sec:Hund_ntot_2} and ~\ref{sec:charge_ntot_2} shows that the temperature of the Hund peak is roughly given by $T_\text{Hund}^{(n=2)}=0.455J$ and that of the charge peak by $T_\text{charge}^{(n=2)}=0.197U$.

Similarly, the model at half-filling ($n=3$) exhibits a Hund excitation peak originating from local spin excitations within the $N_\Gamma=3$ sector, and a charge excitation peak at higher temperature, see panel (f). The estimates for the corresponding peak temperatures are $T_\text{Hund}^{(n=3)}=0.873J$ and $T_\text{charge}^{(n=3)}=0.198U$, see Appendix \ref{sec:Hund_ntot_3} and \ref{sec:charge_ntot_3}. 

In Table~\ref{tab_peaks} we list $T_\text{Hund}$ and $T_\text{charge}$ obtained from the atomic model analysis. We note that while the position of the Hund peak is proportional to $J$, as expected, there are nontrivial prefactors. Also the result for the charge peak is very different from naive estimates of the charge gap (such as $U-W$, with $W$ the bandwidth).  

\begin{table}
\caption{\label{tab:Atomic-Hund-charge-peak}
Positions of the Hund peak and charge peak in the electronic specific heat $C_V(T)$ of
the three-orbital atom. 
}
\begin{ruledtabular}
\begin{tabular}{cccc}
Filling &  $n$ & Hund Peak & Charge Peak\tabularnewline
\hline 
$1/6$ & 1 &  & $0.201U$\tabularnewline
$1/3$ & 2 & $0.455J$ & $0.197U$\tabularnewline
$1/2$ & 3 & $0.873J$ & $0.198U$\tabularnewline
\end{tabular}
\end{ruledtabular}
\label{tab_peaks}
\end{table}

The results for model (i) show qualitatively similar features, but in this case $J$ is not much smaller than $U$ and for a quantitatively correct estimate of the charge peak, one needs to take into account the effect of $J$ on the energies of the individual states in sectors $N_\Gamma=2$, $3$ and $4$. For example, in the model with $n=1$ (panel (a)), a more accurate estimate of the charge peak is $T_\text{charge}^{(n=1)}=0.214(U-3J)$, see Appendix~\ref{sec:Cv_n_eq_1_Ueff}. In the model with $n=2$, the probability distribution reveals that the low-energy peak in the specific heat near $T=0.1$ originates from both charge and spin excitations, while the higher energy peak comes from charge excitations to the higher-energy (i.e. low-spin) states within the $N_\Gamma=3$ sector. Also in the model with $n=3$ the spin excitations within the $N_\Gamma=3$ sector are activated at roughly the same temperature as the charge excitations to the neighboring $N_\Gamma=2$ and 4 sectors, which results in a single broad $C_V(T)$ peak near $T=0.4$. Because of the smaller $U$ the $N_\Gamma=1$ and 4 sectors get populated at $T\gtrsim 1$, which leads to a broadening of the peak. 

Figure~\ref{cv_atomic_full_and_subspace_ctqmc_n1n2n3} illustrates how well the atomic model analysis allows to explain the features in the specific heat of the lattice model. The red curves in the figure show the DMFT results for integer fillings (top panels for model (i) and bottom panels for model (ii)). These results are compared to the atomic model solution which considers the dominant sector $N_\Gamma=n$ and the two neighboring sectors $n-1$ and $n+1$ (yellow), and to an atomic model calculation which takes into account only spin excitations (green).
The dashed blue line, obtained by setting $J=0$, represents the contribution from charge excitations. 
It is found that in model (ii), which is Mott insulating at $n=$1, 2, and 3, the atomic model analysis almost perfectly explains the origin of the peaks in the specific heat. On the other hand, in model (i), which is metallic for $n=1$ and $2$  and has a small gap at $n=3$ there is no quantitative agreement between the atomic model and DMFT data. In particular, the specific heat in the metallic systems is substantially larger than predicted by the atomic model. Still, the Hund and charge peaks identified in the atomic model explain the positions and widths of the peaks in $C_V(T)$.

This is also confirmed in the lower panels of Fig.~\ref{EntropyFixn_T}, where the positions of the atomic spin (charge) peaks are indicated by thin (thick) dashed lines. Here, we have used the values in Tab.~\ref{tab_peaks}, except for the blue dashed line in panel (b), which shows more accurate estimate $T_\text{charge}^{(n=1)}=0.214(U-3J)$ (see Appendix \ref{sec:Cv_n_eq_1_Ueff}).

Finally, let us refer to the third row in Fig.~\ref{schematic}, which shows the specific heat of the noninteracting model, model (i), and model (ii) for arbitrary fillings. In model (i), which is Mott insulating only at $n=3$, we recognize at $T\approx 0.4$ the combined spin and charge excitation peak of Fig.~\ref{atomic_prob_and_Cv}(c) at half-filling, 
which persists as filling is reduced and merges into the higher energy charge excitation feature evident in Fig.~\ref{atomic_prob_and_Cv}(b). The lower energy spin/charge peak of the $n=2$ model, on the other hand, is continuously connected to the charge excitation feature at $n=1$, and down to lower fillings, where it becomes hardly distinguishable from the peak in the noninteracting model. In the results for model (ii), we recognize the spin peak at $n=3$, the weak spin peak at $n=2$ and the completely absent spin peak at $n=1$, while the charge peaks are similarly prominent at all the fillings, but do not extend very far into the doped metal regimes. On the other hand, the spin peak persists in the filling range $2\le n \le 3$, which is consistent with our previous remark that this entire filling region should at low (but nonzero) temperatures be regarded as a spin-frozen region associated with the $n=3$ Mott insulator.   

\section{Discussion and Conclusions}
\label{sec:discussion}

We have calculated the entropy and specific heat of the three-orbital Hubbard model on the infinitely connected Bethe lattice. At moderate $U$, where only the half-filled solution is Mott insulating, the Hund's coupling induces pronounced spin-freezing and non-Fermi-liquid effects in a wide doping and temperature range. The entropy and specific heat provide an interesting perspective on this Hund metal behavior. In particular, we showed that the entropy per site in the spin-frozen metal is enhanced ($S \gtrsim \ln 2$) and smoothly connected to the $\ln 2$ entropy of the half-filled Mott state. The crossover to the Fermi liquid state at low temperatures is associated with a screening of the local moments and hence with a rapid drop of the entropy per site from $\sim \ln 2$ to 0.  

The crossover to the spin-frozen region is associated with a maximum in the dynamical contribution to the local spin susceptibility, $\Delta\chi_\text{loc}$, as previously suggested in Ref.~\onlinecite{Hoshino2015}. The corresponding crossover line is meaningful in particular at low temperatures, where the peak in $\Delta\chi_\text{loc}$ is pronounced. We showed that above the characteristic temperature scale for the activation of local spin excitations, the peak amplitude drops quickly so that the spin-freezing crossover loses its significance. This is natural, because the thermal population of different local spin states washes out the spin-freezing effect.  

The activation of spin and charge fluctuations is associated with peaks in the specific heat. We have analyzed these peaks in the atomic model with integer fillings, and showed that these results provide a qualitative understanding of $C_V(n,T)$ for arbitrary fillings $n$. In particular, we found that the temperature scale $T_\text{spin}$ for Hund excitations is determined only by $J$ (as expected), but with a nontrivial prefactor that depends on the filling. We also showed that for large enough $U$ and small enough $J/U$, the charge excitation peak in the specific heat occurs at $T_\text{charge}\approx U/5$. Especially at large $U$, this is much smaller than naive estimates of the charge gap, which shows that in the three-orbital model, in contrast to the single-orbital Hubbard model, the Mott gap is not very robust against temperature.  

In our model (i) with $U=2$, $J=0.5$ the spin and charge excitations are activated at a comparable temperature, so that there is only a single high-temperature crossover in the entropy from the low-temperature Fermi liquid or the $\ln 2$ 
plateau to the infinite temperature value of $-6[\frac{n}{6}\ln\frac{n}{6}+(1-\frac{n}{6})\ln(1-\frac{n}{6})].$ In model (ii) with $U=8$, $J=0.25$, $T_\text{spin}$ and $T_\text{charge}$ are clearly separated, so that the entropy exhibits an intermediate plateau corresponding to a system with thermally activated local spin excitations but still suppressed charge excitations. The spin freezing behavior of model (ii) is different from what has been previously discussed in the Hund metal literature, because this model is Mott insulating at $n=1$, $2$ and $3$. The system hence exhibits local moment formation near all these integer fillings. Nevertheless, the evolution of the local spin susceptibility and of $\Delta\chi_\text{loc}$ show that the spin-freezing associated with the half-filled Mott state affects the metallic solutions down to fillings below $n=2$, similar to the case of model (i). 

In future studies, it would be very interesting to complement the present picture of the entropy and specific heat in the temperature range $0.005 \le T \le \infty$ with a systematic analysis of the very low temperature behavior near and in the spin-frozen region. Because of the very low Fermi liquid coherence scale near half-filling, the crossover from the high-entropy spin-frozen metal with $S\gtrsim \ln 2$ to the Fermi-liquid metal with $S\propto T$ cannot be studied with CT-HYB and likely requires an NRG-based investigation \cite{STADLER2019365}. It will also be interesting to extend the current study to models with $J<0$, relevant for fulleride compounds. The orbital-freezing crossovers and spontaneous orbital selective Mott phases \cite{Hoshino2017} in these models should lead to nontrivial structures in the specific heat and entropy surfaces. 

\acknowledgments{
The calculations have been performed on the Beo05 cluster at the University of Fribourg and the Tianhe-1A platform at the National Supercomputer Center in Tianjin, 
using a code based on iQist \cite{HUANG2015140,iqist}. We acknowledge support from SNSF Grant No.~200021-165539. 
}

\begin{widetext}
\appendix
\numberwithin{equation}{section}
\numberwithin{figure}{section}

\section{Hubbard-I approximation} 
\label{app_HubbardI}
At very high temperature, $T\gg t$, the correlated system approaches the atomic limit where the local Green's function can be well 
described by the Hubbard-I approximation \cite{HubI_1964} 
\begin{align}
G_{\alpha}(i\omega_{n})\approx G^{\text{Hub-I}}_{\alpha}(i\omega_{n}) & =\sum_{\Gamma\Gamma^{\prime}}\frac{(\frac{1}{Z}e^{-\beta E_{\Gamma}}+\frac{1}{Z}e^{-\beta E_{\Gamma^{\prime}}})|\langle\Gamma|d_{\alpha}|\Gamma^{\prime}\rangle|^{2}}{i\omega_{n}-(E_{\Gamma^{\prime}}-E_{\Gamma})},
\label{eq:GiwHubI}
\end{align}
where $Z=\sum_{\Gamma}e^{-\beta E_{\Gamma}}$ is the atomic partition function, and $E_{\Gamma}$ the eigenvalue of $H_{\text{loc}}$ in the 
eigenstate $\Gamma$. The imaginary-time Hubbard-I Green's function reads 
\begin{equation}
G_{\alpha}^{\text{Hub-I}}(\tau)=-\sum_{\Gamma}p_{\Gamma}\sum_{\Gamma^{\prime}}e^{-\tau(E_{\Gamma^{\prime}}-E_{\Gamma})}|\langle\Gamma|d_{\alpha}|\Gamma^{\prime}\rangle|^{2},
\end{equation}
where $p_\Gamma \equiv e^{-\beta E_{\Gamma}}/Z$ denotes the probability of the eigenstate $\Gamma$ and $0\le \tau \le \beta$.
The energy in the Hubbard-I approximation is 
 \cite{prb2007_Haule_qmc,prb2007_Werner_dopeMott} 
\begin{equation}
E_{\text{tot}}(n,T) = t^2 \sum_{\alpha}\int_{0}^{\beta} d\tau G^{\text{Hub-I}}_{\alpha}(\tau) G^{\text{Hub-I}}_{\alpha}(-\tau)+\sum_{\Gamma}p_{\Gamma}E_{\Gamma}, 
\end{equation}
where the first term represents the kinetic energy for a Bethe lattice and the second term the local energy. A comparison between the numerically exact total energy sampled by CT-HYB and that calculated with 
the Hubbard-I approximation is shown in Fig.~\ref{etot_hub1_ctqmc}. At $T=1.0$, there is still a slight discrepancy between these two results 
in the $U=2.0, J=0.50$ system while the difference is already negligible in the $U=8.0, J=0.25$ model which is closer to the 
atomic limit.  At the higher temperature $T=4.0$, the CT-HYB total energy can be well approximated by the Hubbard-I energy in both cases.

\begin{figure*}[t]
\includegraphics[clip,width=0.8\paperwidth,angle=0]{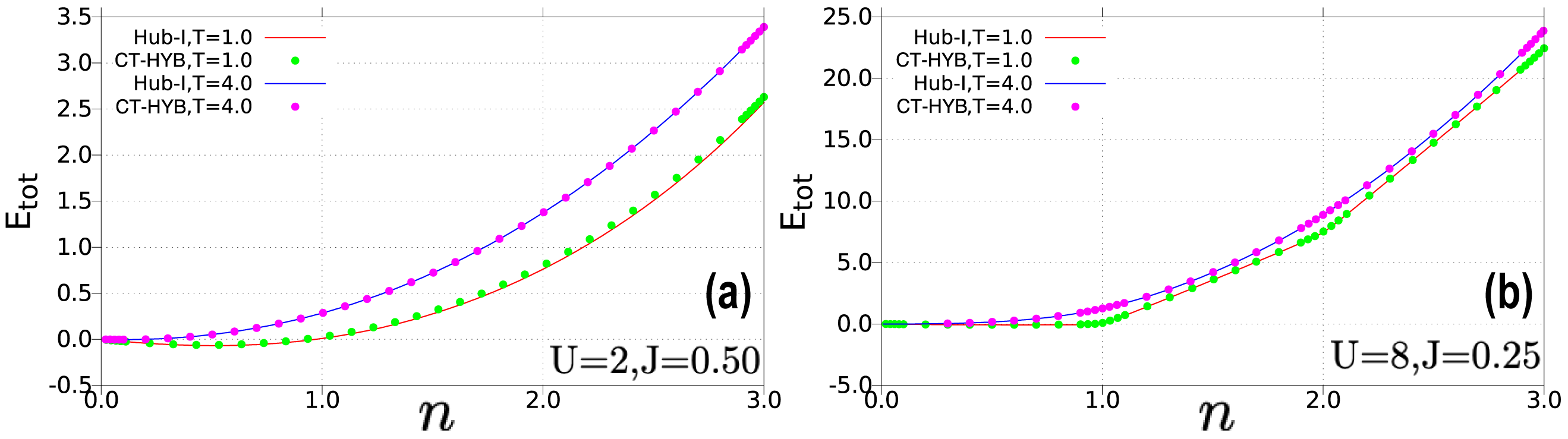}
\caption{ (color online). Filling dependence of  the total energy measured by CT-HYB and in the Hubbard-I 
approximation at the high temperatures $T=1.0$ and $T=4.0$. Panel (a) is for $U=2.0, J=0.50$ and panel (b) for $U=8.0, J=0.25$. 
}
\label{etot_hub1_ctqmc}
\end{figure*}

\section{Virtual updates for measuring the Green's function at high temperatures}\label{sec:virtual_update}

The conventional estimator \cite{Lewin_prb_2011_polynomial} for the one-particle Green's function
reads,
\begin{equation}
G_{\alpha\alpha}(\tau)=-\frac{1}{\beta}\left\langle \sum_{nm}^{k_{\alpha}}\text{sgn}\frac{\text{det}\boldsymbol{\Delta}_{\alpha}^{(nm)}}{\text{det}\boldsymbol{\Delta}_{\alpha}}\delta\left(\tau,\tau_{m}^\alpha-\tau_{n}'^\alpha\right)
\right\rangle _{\mathrm{MC}},\label{eq:G_Estimator_old}
\end{equation}
where for simplicity we 
consider the case where the hybridization matrix $\boldsymbol{\Delta}$ and the Green's function matrix 
are diagonal in the flavors (spin-orbitals) $\{\alpha\}$. 
Here $\boldsymbol{\Delta}_{\alpha}^{(nm)}$ represents the matrix with the row and column corresponding to the $m$th annihilation operator and $n$th creation operator removed.  
We can separate the contributions to $G_{\alpha\alpha}(\tau)$ in Eq.~(\ref{eq:G_Estimator_old})
into contributions from the different expansion orders $k_{\alpha}$: 
\begin{equation}
G_{\alpha\alpha}(\tau)=\sum_{k_{\alpha}=0}^{\infty}\langle G_{\alpha\alpha}^{(k_{\alpha})}(\tau)\rangle,
\end{equation}
where 
\begin{equation}
G_{\alpha\alpha}^{(k_{\alpha})}(\tau)=-\frac{1}{\beta}\sum_{nm}^{k_{\alpha}}\text{sgn}\frac{\text{det}\boldsymbol{\Delta}_{\alpha}^{(nm)}}{\text{det}\boldsymbol{\Delta}_{\alpha}}\delta\left(\tau,\tau_{m}^\alpha-\tau_{n}'^\alpha \right). 
\end{equation}
The conventional estimator is based on the removal of hybridization lines between
pairs of creation and annihilation operators in a $k_{\alpha}$th order diagram
of the partition function, so that
the $0$th order configurations contribute zero 
to $G_{\alpha\alpha}(\tau)$. 
As a consequence, $G_{\alpha\alpha}(\tau)$ becomes 
noisy when the $0$th order diagrams of the partition function are dominant,
i.e. the average expansion order $\langle k_{\alpha}\rangle\rightarrow0$, 
since higher-order diagrams are rarely generated. This problem can become serious in the large $U$ limit, 
weak hybridization limit, or in the high-temperature limit. To reduce the noise, one
may resort to a worm sampling algorithm \cite{ctqmc_worm_prb2015}.
Alternatively, this noise problem can be solved by virtual updates between the expansion orders $0$ and $1$.  
The word ``virtual'' here means that such kinds of updates 
are merely performed for the sake of measurement, not in the actual sampling process. 

The procedure is as follows. In the CT-HYB simulations,
the Green's function is measured periodically, after a certain number 
of updates which depends on the auto-correlation time. 
At the time of measurement we perform virtual updates in configurations with perturbation order $0$ and $1$:

\begin{enumerate}
\item If $k_{\alpha}=0$, a virtual hybridization insertion is proposed with 
$d_{\alpha}$ randomly located at $\tau_{1}$ and $d_{\alpha}^{\dagger}$
at $\tau_{2}'$. The new estimator at $k_{\alpha}=0$ reads 
\begin{align*}
G_{0\rightarrow1,\alpha}^{\text{vir}}(\tau) & = \langle p_{0\rightarrow1}^{\text{acc}} G_{\alpha\alpha}^{(k_{\alpha}=1)}(\tau)\rangle+\langle (1-p_{0\rightarrow1}^{\text{acc}}) G_{\alpha\alpha}^{(k_{\alpha}=0)}(\tau)\rangle = \langle p_{0\rightarrow1}^{\text{acc}} G_{\alpha\alpha}^{(k_{\alpha}=1)}(\tau)\rangle,
\end{align*}
where we used that $G_{\alpha\alpha}^{(k_{\alpha}=0)}(\tau)=0$. The proposal probability for a virtual insertion is the same as for a normal
insertion update \cite{ctqmc_prl2006} from $k_{\alpha}=0$ to $k_{\alpha}=1$, and the 
acceptance probability reads 
\begin{equation}
p_{0\rightarrow1}^{\text{acc}}=\min\left\{ 1,\frac{\beta^{2}}{(d\tau)^2}\frac{|p_{1}|}{|p_{0}|}\right\} ,
\end{equation}
with 
\begin{equation}
\frac{p_{1}}{p_{0}}=\frac{(d\tau)^2\text{Tr}_{d}\left[\mathcal{T}_{\tau}e^{-\beta H_{\text{loc}}}d_{\alpha}(\tau_{1})d_{\alpha}^{\dagger}(\tau_{2}')\right]\Delta_{\alpha}(\tau_{1}-\tau_{2})}{\text{Tr}_{d}[e^{-\beta H_{\text{loc}}}]}.
\end{equation}
If the acceptance probability for the virtual move from 0 to 1 is less than one, 
as is typically the case if diagrams with $k_{\alpha}=0$ dominante,
the estimator becomes 
\begin{align*}
\langle p_{0\rightarrow1}^{\text{acc}} G_{\alpha\alpha}^{(k_{\alpha}=1)}(\tau)\rangle 
& = \frac{1}{\beta^2}\int_0^\beta d\tau_1 d\tau_2' \beta^{2} \frac{\text{Tr}_{d}\left[\mathcal{T}_{\tau}e^{-\beta H_{\text{loc}}}d_{\alpha}(\tau_{1})d_{\alpha}^{\dagger}(\tau_{2}')\right]}{\text{Tr}_{d}[e^{-\beta H_{\text{loc}}}]}\Delta_{\alpha}(\tau_{1}-\tau_{2}')\frac{-1}{\beta}\text{\text{sgn}}\frac{1}{\Delta_{\alpha}(\tau_{1}-\tau_{2}')}\delta\left(\tau,\tau_{1}-\tau_{2}'\right)\\
&=-\frac{\text{Tr}_{d}\left[\mathcal{T}_{\tau}e^{-\beta H_{\text{loc}}}d_{\alpha}(\tau)d_{\alpha}^{\dagger}(0)\right]}{\text{Tr}_{d}[e^{-\beta H_{\text{loc}}}]} = G^\text{atom}_{\alpha\alpha}(\tau).
\end{align*}
If, on the other hand, the acceptance rate is 1, we measure $G_{\alpha\alpha}^{(k_\alpha=1)}(\tau)$ for the corresponding virtual configuration with $k_\alpha=1$.
The sampling with virtual updates thus accumulates the following Green's function estimator at $k_{\alpha}=0$:
\begin{equation}
G^\text{atom}_{\alpha\alpha}(\tau) \text{  if  }p_{0\rightarrow1}^{\text{acc}}<1, \quad G_{\alpha\alpha}^{(k_\alpha=1)}(\tau) \text{  otherwise}.
\end{equation}

\begin{figure*}[b]
\includegraphics[clip,width=0.8\paperwidth,angle=0]{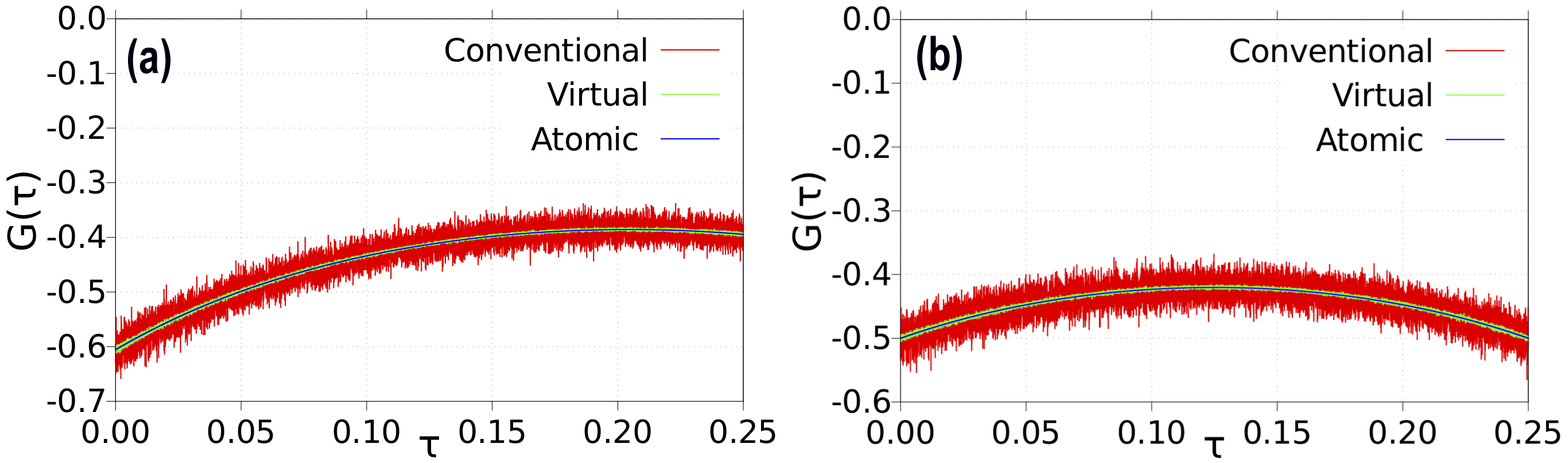}
\caption{ (color online). Comparison of the imaginary-time Green's functions obtained by the conventional estimator and virtual updates estimator to the atomic Green's function for $U=8$, $J=0.25$, $T=4.0$. Panel (a) is for chemical potential $\mu=12.5$ and (b) is for 
$\mu=18.75$. The results are for a one-shot calculation with a symmetric, noninteracting hybridization function as input.}
\label{gtau_compareU8J0.25T4}
\end{figure*}

\item If $k_{\alpha}=1$, one proposes a virtual hybridization line removal update.
The new estimator at $k_{\alpha}=1$ reads 
\begin{align*}
G_{1\rightarrow0,\alpha}^{\text{vir}}(\tau) & =\langle p_{1\rightarrow0}^{\text{acc}}  G_{\alpha\alpha}^{(k_{\alpha}=0)}(\tau)\rangle+\langle (1-p_{1\rightarrow0}^{\text{acc}}) G_{\alpha\alpha}^{(k_{\alpha}=1)}(\tau)\rangle
 =\langle (1-p_{1\rightarrow0}^{\text{acc}}) G_{\alpha\alpha}^{(k_{\alpha}=1)}(\tau)\rangle
\end{align*}
with
\begin{equation}
p_{1\rightarrow0}^{\text{acc}}=\min\left\{ 1,\frac{(d\tau)^2}{\beta^{2}}\frac{|p_{0}|}{|p_{1}|}\right\}.
\end{equation}
This means that we perform the usual measurement in the configuration with $k_\alpha=1$ if the virtual update to the empty configuration is rejected, otherwise we measure $0$.
\item If $k_{\alpha}\ge2$, use the conventional estimator $G_{\alpha\alpha}^{(k_{\alpha})}(\tau)$.
\end{enumerate}
This modified sampling method is exact and automatically yields the atomic Green's function in parameter regimes where the average perturbation order goes to zero. 

In Fig.~\ref{gtau_compareU8J0.25T4}, we compare the Green's
function measured by the conventional estimator (red) and the virtual updates estimator (green)
to the atomic Green's function of the impurity model with parameters $U=8$, $J=0.25$ at high temperature $T=4$. The 
results demonstrate that the new estimator can substantially reduce the noise and allows to access the atomic limit. 

\section{Global Updates}\label{sec:Global_Updates}
When the system is near a Mott state at low temperatures, the Monte Carlo sampling can be trapped in certain configurations. Global updates which exchange the segments of two spin-orbitals have been proposed \cite{rmp_ctqmc}. 
However, such global updates alone are not enough to get rid of trapping. Here we propose additional
global updates. One is the ``double swap" update, 
where occupied segments in two different spin-orbitals are replaced with unoccupied anti-segments, see Fig.~\ref{global_updates_pic}(b)$\rightarrow$(a). 
This is equivalent to exchanging the creation and annihilation operators, without changing the imaginary times and flavors. 
A special case of this update (for perturbation orders $0$) is the exchange of a full line in one flavor with an empty line in another flavor, as shown in
Fig.~\ref{global_updates_pic}(d)$\rightarrow$(e). This update is helpful at integer fillings and at high temperatures, where the expansion order is low. The second global update is a global shift operation
which is similar to the ``global $\tau$ shift update" proposed in Ref.~\onlinecite{state_sampling_prb2019}. 
Here, instead of shifting all the operators by $\tau$, we shift only the operators for a single flavor by $\tau\in(-\beta,+\beta)$. This update is equivalent to $k_{\alpha \sigma}$ removal updates and $k_{\alpha \sigma}$ insertion 
updates where $k_{\alpha \sigma}$ is the expansion order of the selected flavor ${\alpha \sigma}$.

\begin{figure*}[t]
\includegraphics[clip,width=0.8\paperwidth,angle=0]{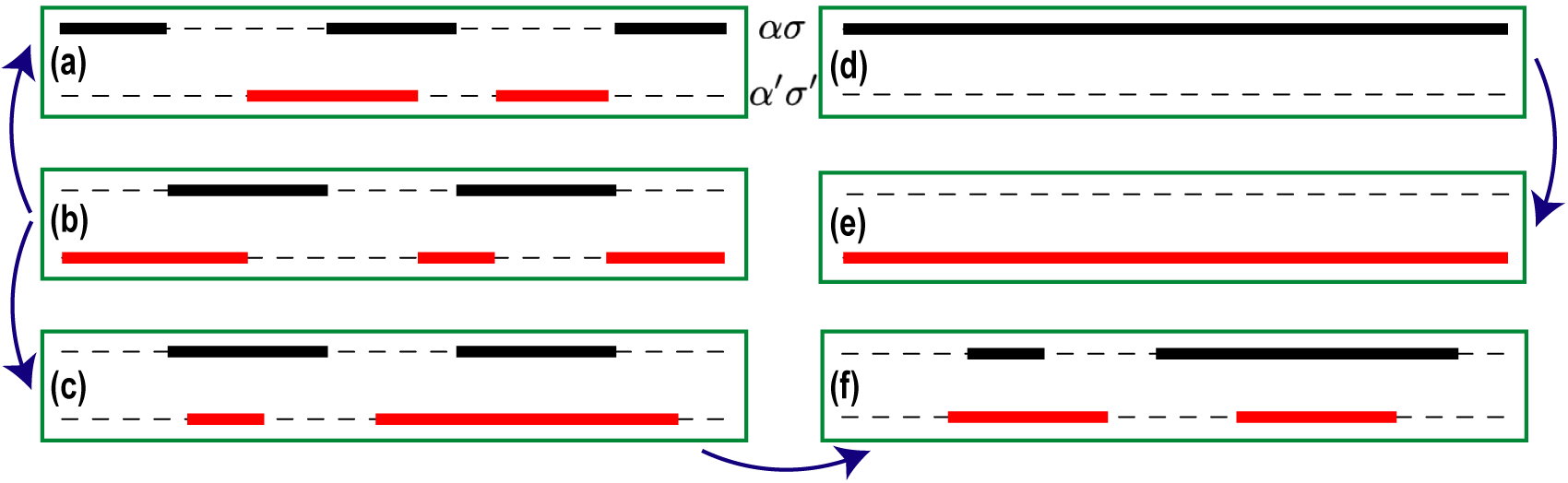}
\caption{ (color online). Global updates of the segment version of CT-HYB. Two different spin-orbitals $\alpha \sigma$ and $\alpha^\prime \sigma^\prime$ are randomly selected for these updates. While the black thin dashed lines represent the unoccupied antisegments, the black and red solid lines denote
the occupied segments in these two spin-orbitals. (c)$\rightarrow $(f) represents an 
update where the segments of two spin-orbitals are exchanged. (b)$\rightarrow$(a) represents a 
double swap update, where segments and anti-segments 
are swapped within two spin-orbitals. 
(d)$\rightarrow$(e) shows an exchange of a full line in one orbital and an empty line in another.
(b)$\rightarrow$(c) represents a single-orbital global shift update where the operators in a randomly
selected orbital (here $\alpha^\prime \sigma^\prime$) are shifted by $\delta \tau \in (-\beta,+\beta)$.
} 
\label{global_updates_pic}
\end{figure*}

\section{Atomic Specific Heat}\label{sec:Atomic-Cv}

The ``Hund peak" and ``charge peak" positions of the atomic electronic specific heat at integer fillings have been collected
in Tab.~\ref{tab:Atomic-Hund-charge-peak} in the main text. In this section, we explain how these results are obtained  using subspace truncations of the atomic Hamiltonian $H_{\text{loc}}$, whose eigenstates are listed in Tab.~\ref{tab:Atomic-Eigenstates}.

\subsection{Charge peak at $n=1$}
\label{sec:charge_peak_ntot_1}
\subsubsection{Estimate for $J=0$}
As illustrated in Fig.~\ref{cv_atomic_full_and_subspace_ctqmc_n1n2n3} (a,d), the charge peak at $n=1$ arises 
from charge fluctuations between the dominant sector 1 and the neighboring sectors 0 and 2. The partition function restricted to these three sectors is
\begin{equation}
Z=6 e^{\frac{2 J+2 \mu-U}{T}}+6 e^{\frac{3 J+2 \mu-U}{T}}+3 e^{\frac{2 \mu-U}{T}}+6 e^{\mu/T}+1.
\end{equation}
To ensure the correct average filling $n=1$, $\mu$ is tuned such that 
\begin{equation}
n=\frac{6 e^{\mu/T} \left(2 e^{\frac{2 J+\mu}{T}}+2 e^{\frac{3 J+\mu}{T}}+e^{\mu/T}+e^{U/T}\right)}{6 e^{\frac{2 (J+\mu)}{T}}+6 e^{\frac{3 J+2 \mu}{T}}+6 e^{\frac{\mu+U}{T}}+3 e^{\frac{2 \mu}{T}}+e^{U/T}}=1,
\end{equation}
which yields 
$\mu=\frac{1}{2} (U-3 J)-\frac{1}{2} T \log (6 e^{-\frac{J}{T}}+3 e^{-\frac{3 J}{T}}+6)$.
The resulting total energy (Eq.~\eqref{eq:E_int_atomic}) is
\begin{equation}
E_{\text{tot}}=\frac{(2 U-6 J) e^{\frac{3 J-U}{T}}+(2 U-4 J) e^{\frac{2 J-U}{T}}+U e^{-\frac{U}{T}}}{2 \sqrt{3} \sqrt{e^{-\frac{3 J}{T}}+2 e^{-\frac{J}{T}}+2} e^{\frac{3 J-U}{2 T}}+4 e^{\frac{2 J-U}{T}}+4 e^{\frac{3 J-U}{T}}+2 e^{-\frac{U}{T}}},
\label{eq:Eint_ntot_1_Jne0}
\end{equation}
which for $J=0$ simplifies to
\begin{equation}
E_{\text{tot}}=\frac{2 U}{2 \sqrt{3} e^{\frac{U}{2 T}}+4}.
\end{equation}
The atomic specific heat is obtained via Eq.~\eqref{Eq_cv}, 
\begin{equation}
C_V(T)=
\frac{2 \sqrt{3} U^2 e^{\frac{U}{2 T}}}{T^2 \left(2 \sqrt{3} e^{\frac{U}{2 T}}+4\right)^2},
\end{equation}
and the 
peak position of $C_V(T)$ is located at the root of 
\begin{equation}
\frac{\partial C_{V}(T)}{\partial T}=-\frac{25 U^2 e^{\frac{U}{2 T}} \left(4 T \left(3 e^{\frac{U}{2 T}}+\sqrt{15}\right)+U \left(\sqrt{15}-3 e^{\frac{U}{2 T}}\right)\right)}{8 T^4 \left(\sqrt{15} e^{\frac{U}{2 T}}+5\right)^3}=0.
\end{equation}
This is equivalent to finding the root of the equation
\begin{equation}
\frac{x}{4}=\frac{3e^{\frac{x}{2}}-\sqrt{15}}{3e^{\frac{x}{2}}+\sqrt{15}},
\label{eq:root_charge_peak_ntot_1}
\end{equation}
with $x\equiv U/T$. The numerical solution of Eq.~\eqref{eq:root_charge_peak_ntot_1} gives $x_0\approx4.966$, so that the charge peak position for $n=1$ is
\begin{equation}
T_{\text{charge}}^{(n=1)}=\frac{1}{x_0} U \approx 0.2014 U.
\label{eq:T_charge_ntot_1}
\end{equation}

\subsection{ More accurate estimate for $n=1$ ($J\ne0$)}
\label{sec:Cv_n_eq_1_Ueff}
According to Eq.~\eqref{eq:T_charge_ntot_1}, the charge peak position for model (ii) ($U=8,J=0.25$) is located 
at 1.61 which is only slightly larger than the value of 1.50 obtained by CT-HYB (see Fig.~\ref{cv_atomic_full_and_subspace_ctqmc_n1n2n3}(d)). However, 
Eq.~\eqref{eq:T_charge_ntot_1} gives $0.40$  for model (i) ($U=2,J=0.50$) which substantially overestimates the CT-HYB peak at $0.11$ (see Fig.~\ref{cv_atomic_full_and_subspace_ctqmc_n1n2n3}(a)). 
For this smaller $U$ and larger $J/U$, the Hund's coupling effect on the charge excitations needs to be taken into account to obtain a good estimate of the charge peak position.

If $J/U$ is large, the relevant sectors are the sectors 0, 1, and the multiplets $\phi_{2G}$ in sector 2, as shown in Fig.~\ref{atomic_prob_and_Cv}(a).
The partition function restricted to these subspaces reads 
\begin{equation}
Z=1+6e^{\frac{\mu}{T}}+6e^{\frac{3J-U+2\mu}{T}}.
\end{equation}
The total filling is 
\begin{align}
n & =\frac{6e^{\frac{\mu}{T}}+12e^{\frac{3J-U+2\mu}{T}}}{1+6e^{\frac{\mu}{T}}+6e^{\frac{3J-U+2\mu}{T}}}
\end{align}
so that the chemical potential $\mu$ corresponding to $n=1$ is given by $\mu=\frac{U-3J}{2}-\frac{\ln6}{2}T$.
The total energy becomes 
\begin{align}
E_{\text{tot}}(T) & =(U-3J)\left(2+\sqrt{6}e^{\frac{U-3J}{2T}}\right)^{-1}, 
\end{align}
and the specific heat 
\begin{equation}
C_{V}(T)= \frac{\sqrt{\frac{3}{2}}(U-3J)^{2}e^{\frac{U-3J}{2T}}}{T^{2}\left(2+\sqrt{6}e^{\frac{U-3J}{2T}}\right)^{2}}.
\label{eq:CvT_ntot_1_Ueff}
\end{equation}
The peak position in the specific heat is determined by requiring that
\begin{equation}
\frac{\partial C_{V}(T)}{\partial T}=\frac{(U-3J)^{2}e^{\frac{U-3J}{2T}}\left(3e^{\frac{U-3J}{2T}}(U-3J-4T)-\sqrt{6}(U-3J+4T)\right)}{2T^{4}\left(2+\sqrt{6}e^{\frac{U-3J}{2T}}\right)^{3}}=0,
\end{equation}
which is equivalent to finding the root of the equation
\begin{equation}
\sqrt{\frac{3}{2}}e^{\frac{x}{2}}=\frac{x+4}{x-4},
\end{equation}
with $x\equiv\frac{U-3J}{T}$. The numerical solution is $x_{0}\approx4.6821$,
so that the more accurate estimate for the charge peak position at $n=1$ is
\begin{equation}
T_{\text{charge}}^{(n=1)}\approx\frac{U-3J}{x_{0}}\approx0.2136(U-3J).
\label{eq:Cv_peak_ntot_1}
\end{equation}

According to Eq.~(\ref{eq:Cv_peak_ntot_1}) the peak position of the atomic specific heat at $n=1$
is controlled by $U-3J$. 
We find $T_{U=2,J=0.5}^{(n=1)}\approx0.107$ and $T_{U=8,J=0.25}^{(n=1)}\approx1.548$,
respectively, which agrees well  with the peak positions in $C_{V}(T)$ obtained by
CT-HYB (see Fig.~\ref{cv_atomic_full_and_subspace_ctqmc_n1n2n3}(a,d)).
Note that Eq.~\eqref{eq:CvT_ntot_1_Ueff} underestimates the peak height of $C_V(T)$ measured by CT-HYB in model (i) since 
the kinetic contribution is important when $U$ is comparable to the band width $W$ (here $U=W=2$).

\subsection{Charge Peak at $n=2$}
\label{sec:charge_ntot_2}
For the charge peak at $n=2$, we need to take into account sectors 1, 2 and 3.
The partition function reads 
\begin{align}
Z & =6e^{\mu/T}+e^{-(U-2\mu)/T}[6e^{3J/T}+6e^{2J/T}+3]+e^{-(3U-3\mu)/T}[2e^{9J/T}+6e^{7J/T}+12e^{5J/T}].
\end{align}
The chemical potential is adjusted to ensure the proper filling,
$\mu =\frac{3U-9J}{2}-\frac{1}{2}T\ln\frac{1+3e^{\frac{-2J}{T}}+6e^{\frac{-4J}{T}}}{3}$.
Considering again for simplicity the case $J=0$, we find that $\mu$ depends linearly on $T$,
$\mu = \frac{3}{2} U -\frac{1}{2} \ln{\frac{10}{3}} \cdot T $,
and the total energy calculated within the $n=1,2,3$ subspaces becomes
\begin{equation}
E_{\text{tot}}(T)=\frac{U\left(\sqrt{30}e^{\frac{U}{2T}}+12\right)}{\sqrt{30}e^{\frac{U}{2T}}+8}.
\end{equation}
The specific heat is 
\begin{equation}
C_{V}(T)=\frac{2\sqrt{30}(\frac{U}{T})^{2}e^{\frac{U}{2T}}}{\left(\sqrt{30}e^{\frac{U}{2T}}+8\right)^{2}},
\end{equation}
and its peak position is located at the root of
\begin{equation}
60e^{x/2}-15xe^{x/2}+16\sqrt{30}+4\sqrt{30}x=0,
\end{equation}
with $x=U/T$. The root value is $x=5.056$, so that the characteristic temperature for 
charge excitations at $n=2$ becomes 
\begin{equation}
T_{\text{charge}}^{(n=2)} = U/5.056 \approx 0.1978 U. 
\end{equation}

\subsection{Charge Peak at $n=3$}
\label{sec:charge_ntot_3}
According to Fig.~\ref{atomic_prob_and_Cv}(c,f), the broad peak of $C_V(T)$ at higher $T$ at $n=3$ 
is related to the charge excitations from sector 3 to sectors 2 and 4.
The atomic partition function restricted to these sectors reads
\begin{align}
Z&=e^{\frac{9}{2}U/T}\left[2e^{-6J/T}+6e^{-8J/T}+12e^{-10J/T}\right] + e^{4U/T}\left[12e^{-7J/T}+12e^{-8J/T}+6e^{-10J/T}\right]
\end{align}
where the chemical potential $\mu=5U/2-5J$ has been used. 
For $J=0$, the total energy becomes
\begin{equation} 
E_{\text{tot}}(T)= \frac{3 U \left(4 e^{\frac{U}{2 T}}+7\right)}{4 e^{\frac{U}{2 T}}+6},
\end{equation} 
and the specific heat reads 
\begin{equation}
C_V(T) = \frac{3 (\frac{U}{T})^2 e^{\frac{U}{2 T}}}{2 \left(2 e^{\frac{U}{2 T}}+3\right)^2}.
\end{equation} 
Its peak position is located at the root of 
\begin{equation}
x=\frac{2e^{x}+3e^{-x}}{2e^{x}-3e^{-x}}
\end{equation}
with $x=U/4T$. The numerical solution of this equation gives $x\approx1.269$, from which we obtain the characteristic temperature for 
charge excitations at $n=3$,
\begin{equation}
T_{\text{charge}}^{(n=3)} = U/(4\cdot 1.269) \approx 0.1970 U. 
\end{equation}

\subsection{Hund peak at $n=2$}
\label{sec:Hund_ntot_2}
As shown in Fig.~\ref{atomic_prob_and_Cv}(b,e), excitations within sector 2 define the Hund's peak at $n=2$.
The lowest energy multiplets $\phi_{2G}$ are 6-fold degenerate high spin states. 
The first excited states $\phi_{21}$ are also 6-fold degenerate but with anti-parallel spins in different orbitals. This spin-flip costs an energy $J$ compared 
with the energy of $\phi_{2G}$. The second excited states $\phi_{22}$ are 3-fold degenerate and 
doubly occupied in one orbital. The excitation from $\phi_{21}$ to $\phi_{22}$ costs another energy $2J$.

The atomic partition function truncated to these subspaces of sector 2 reads
\begin{equation}
\begin{aligned}
Z & =e^{-(U-2\mu)/T}[6e^{3J/T}+6e^{2J/T}+3]
\end{aligned}
\end{equation}
and the total energy becomes 
\begin{equation}
\begin{aligned}
E_{\text{tot}}(T) & =\frac{6(U-3J)+6(U-2J)e^{-J/T}+3Ue^{-3J/T}}{6+6e^{-J/T}+3e^{-3J/T}}.
\end{aligned}
\end{equation}
Note that the chemical potential term drops out. The specific heat is
\begin{equation}
C_{V}(T) =\ensuremath{\frac{2{e}^{\frac{2{J}}{{T}}}\left(4+9{e}^{{J}/{T}}+2{e}^{\frac{3{J}}{{T}}}\right){J}^{2}}{(1+2e^{\frac{2{J}}{{T}}}+2{e}^{\frac{3{J}}{{T}}})^{2}{T}^{2}}},
\end{equation}
which is a function of $J/T$ only. 
The peak position of $C_{V}(T)$ is located at
\begin{equation}
T_{\text{Hund}}^{(n=2)}\approx0.455J.
\label{T_Hund_2}
\end{equation}

\subsection{Hund peak at $n=3$}
\label{sec:Hund_ntot_3}
The active subspace for the Hund peak at $n=3$ is sector 3. The atomic partition function at low temperatures can thus be approximated by
\begin{equation}
Z = e^{(3\mu-3U)/T}\left[2e^{9J/T}+6e^{7J/T}+12e^{5J/T}\right].
\end{equation}
and the total energy by
\begin{equation}
\begin{aligned}
E_{\text{tot}}(T) & =\frac{2(3U-9J)e^{9J/T}+6(3U-7J)e^{7J/T}+12(3U-5J)e^{5J/T}}{2e^{9J/T}+6e^{7J/T}+12e^{5J/T}},
\end{aligned}
\end{equation}
where the chemical potential term also drops out. The specific becomes
\begin{equation}
C_{V}(T)=\frac{12J^{2}e^{\frac{2J}{T}}\left(8e^{\frac{2J}{T}}+e^{\frac{4J}{T}}+6\right)}{T^{2}\left(3e^{\frac{2J}{T}}+e^{\frac{4J}{T}}+6\right)^{2}},\label{HundPea_Cv_n3}
\end{equation}
and is again a function of $J/T$. The peak position of $C_{V}(T)$ is located at
\begin{equation}
T_{\text{Hund}}^{(n=3)}\approx0.873J.
\end{equation}

\end{widetext}

\clearpage
\bibliography{EntropySpinFreezing.bib}

\begin{thebibliography}{61}%
\makeatletter
\providecommand \@ifxundefined [1]{%
 \@ifx{#1\undefined}
}%
\providecommand \@ifnum [1]{%
 \ifnum #1\expandafter \@firstoftwo
 \else \expandafter \@secondoftwo
 \fi
}%
\providecommand \@ifx [1]{%
 \ifx #1\expandafter \@firstoftwo
 \else \expandafter \@secondoftwo
 \fi
}%
\providecommand \natexlab [1]{#1}%
\providecommand \enquote  [1]{``#1''}%
\providecommand \bibnamefont  [1]{#1}%
\providecommand \bibfnamefont [1]{#1}%
\providecommand \citenamefont [1]{#1}%
\providecommand \href@noop [0]{\@secondoftwo}%
\providecommand \href [0]{\begingroup \@sanitize@url \@href}%
\providecommand \@href[1]{\@@startlink{#1}\@@href}%
\providecommand \@@href[1]{\endgroup#1\@@endlink}%
\providecommand \@sanitize@url [0]{\catcode `\\12\catcode `\$12\catcode
  `\&12\catcode `\#12\catcode `\^12\catcode `\_12\catcode `\%12\relax}%
\providecommand \@@startlink[1]{}%
\providecommand \@@endlink[0]{}%
\providecommand \url  [0]{\begingroup\@sanitize@url \@url }%
\providecommand \@url [1]{\endgroup\@href {#1}{\urlprefix }}%
\providecommand \urlprefix  [0]{URL }%
\providecommand \Eprint [0]{\href }%
\providecommand \doibase [0]{http://dx.doi.org/}%
\providecommand \selectlanguage [0]{\@gobble}%
\providecommand \bibinfo  [0]{\@secondoftwo}%
\providecommand \bibfield  [0]{\@secondoftwo}%
\providecommand \translation [1]{[#1]}%
\providecommand \BibitemOpen [0]{}%
\providecommand \bibitemStop [0]{}%
\providecommand \bibitemNoStop [0]{.\EOS\space}%
\providecommand \EOS [0]{\spacefactor3000\relax}%
\providecommand \BibitemShut  [1]{\csname bibitem#1\endcsname}%
\let\auto@bib@innerbib\@empty
\bibitem [{\citenamefont {Georges}\ \emph {et~al.}(2013)\citenamefont
  {Georges}, \citenamefont {Medici},\ and\ \citenamefont
  {Mravlje}}]{annurev_Hund_2013}%
  \BibitemOpen
  \bibfield  {author} {\bibinfo {author} {\bibfnamefont {A.}~\bibnamefont
  {Georges}}, \bibinfo {author} {\bibfnamefont {L.~d.}\ \bibnamefont {Medici}},
  \ and\ \bibinfo {author} {\bibfnamefont {J.}~\bibnamefont {Mravlje}},\ }\href
  {\doibase 10.1146/annurev-conmatphys-020911-125045} {\bibfield  {journal}
  {\bibinfo  {journal} {Annual Review of Condensed Matter Physics}\ }\textbf
  {\bibinfo {volume} {4}},\ \bibinfo {pages} {137} (\bibinfo {year} {2013})},\
  \Eprint
  {http://arxiv.org/abs/https://doi.org/10.1146/annurev-conmatphys-020911-125045}
  {https://doi.org/10.1146/annurev-conmatphys-020911-125045} \BibitemShut
  {NoStop}%
\bibitem [{\citenamefont {Pavarini}\ \emph {et~al.}(2004)\citenamefont
  {Pavarini}, \citenamefont {Biermann}, \citenamefont {Poteryaev},
  \citenamefont {Lichtenstein}, \citenamefont {Georges},\ and\ \citenamefont
  {Andersen}}]{PhysRevLett.92.176403}%
  \BibitemOpen
  \bibfield  {author} {\bibinfo {author} {\bibfnamefont {E.}~\bibnamefont
  {Pavarini}}, \bibinfo {author} {\bibfnamefont {S.}~\bibnamefont {Biermann}},
  \bibinfo {author} {\bibfnamefont {A.}~\bibnamefont {Poteryaev}}, \bibinfo
  {author} {\bibfnamefont {A.~I.}\ \bibnamefont {Lichtenstein}}, \bibinfo
  {author} {\bibfnamefont {A.}~\bibnamefont {Georges}}, \ and\ \bibinfo
  {author} {\bibfnamefont {O.~K.}\ \bibnamefont {Andersen}},\ }\href {\doibase
  10.1103/PhysRevLett.92.176403} {\bibfield  {journal} {\bibinfo  {journal}
  {Phys. Rev. Lett.}\ }\textbf {\bibinfo {volume} {92}},\ \bibinfo {pages}
  {176403} (\bibinfo {year} {2004})}\BibitemShut {NoStop}%
\bibitem [{\citenamefont {Hebard}\ \emph {et~al.}(1991)\citenamefont {Hebard},
  \citenamefont {Rosseinsky}, \citenamefont {Haddon}, \citenamefont {Murphy},
  \citenamefont {Glarum}, \citenamefont {Palstra}, \citenamefont {Ramirez},\
  and\ \citenamefont {Kortan}}]{hebard1991}%
  \BibitemOpen
  \bibfield  {author} {\bibinfo {author} {\bibfnamefont {A.~F.}\ \bibnamefont
  {Hebard}}, \bibinfo {author} {\bibfnamefont {M.~J.}\ \bibnamefont
  {Rosseinsky}}, \bibinfo {author} {\bibfnamefont {R.~C.}\ \bibnamefont
  {Haddon}}, \bibinfo {author} {\bibfnamefont {D.~W.}\ \bibnamefont {Murphy}},
  \bibinfo {author} {\bibfnamefont {S.~H.}\ \bibnamefont {Glarum}}, \bibinfo
  {author} {\bibfnamefont {T.~T.~M.}\ \bibnamefont {Palstra}}, \bibinfo
  {author} {\bibfnamefont {A.~P.}\ \bibnamefont {Ramirez}}, \ and\ \bibinfo
  {author} {\bibfnamefont {A.~R.}\ \bibnamefont {Kortan}},\ }\href {\doibase
  10.1038/350600a0} {\bibfield  {journal} {\bibinfo  {journal} {Nature}\
  }\textbf {\bibinfo {volume} {350}},\ \bibinfo {pages} {600} (\bibinfo {year}
  {1991})}\BibitemShut {NoStop}%
\bibitem [{\citenamefont {Rosseinsky}\ \emph {et~al.}(1991)\citenamefont
  {Rosseinsky}, \citenamefont {Ramirez}, \citenamefont {Glarum}, \citenamefont
  {Murphy}, \citenamefont {Haddon}, \citenamefont {Hebard}, \citenamefont
  {Palstra}, \citenamefont {Kortan}, \citenamefont {Zahurak},\ and\
  \citenamefont {Makhija}}]{rosseinsky1991}%
  \BibitemOpen
  \bibfield  {author} {\bibinfo {author} {\bibfnamefont {M.~J.}\ \bibnamefont
  {Rosseinsky}}, \bibinfo {author} {\bibfnamefont {A.~P.}\ \bibnamefont
  {Ramirez}}, \bibinfo {author} {\bibfnamefont {S.~H.}\ \bibnamefont {Glarum}},
  \bibinfo {author} {\bibfnamefont {D.~W.}\ \bibnamefont {Murphy}}, \bibinfo
  {author} {\bibfnamefont {R.~C.}\ \bibnamefont {Haddon}}, \bibinfo {author}
  {\bibfnamefont {A.~F.}\ \bibnamefont {Hebard}}, \bibinfo {author}
  {\bibfnamefont {T.~T.~M.}\ \bibnamefont {Palstra}}, \bibinfo {author}
  {\bibfnamefont {A.~R.}\ \bibnamefont {Kortan}}, \bibinfo {author}
  {\bibfnamefont {S.~M.}\ \bibnamefont {Zahurak}}, \ and\ \bibinfo {author}
  {\bibfnamefont {A.~V.}\ \bibnamefont {Makhija}},\ }\href {\doibase
  10.1103/PhysRevLett.66.2830} {\bibfield  {journal} {\bibinfo  {journal}
  {Phys. Rev. Lett.}\ }\textbf {\bibinfo {volume} {66}},\ \bibinfo {pages}
  {2830} (\bibinfo {year} {1991})}\BibitemShut {NoStop}%
\bibitem [{\citenamefont {Capone}\ \emph {et~al.}(2009)\citenamefont {Capone},
  \citenamefont {Fabrizio}, \citenamefont {Castellani},\ and\ \citenamefont
  {Tosatti}}]{RevModPhys.81.943}%
  \BibitemOpen
  \bibfield  {author} {\bibinfo {author} {\bibfnamefont {M.}~\bibnamefont
  {Capone}}, \bibinfo {author} {\bibfnamefont {M.}~\bibnamefont {Fabrizio}},
  \bibinfo {author} {\bibfnamefont {C.}~\bibnamefont {Castellani}}, \ and\
  \bibinfo {author} {\bibfnamefont {E.}~\bibnamefont {Tosatti}},\ }\href
  {\doibase 10.1103/RevModPhys.81.943} {\bibfield  {journal} {\bibinfo
  {journal} {Rev. Mod. Phys.}\ }\textbf {\bibinfo {volume} {81}},\ \bibinfo
  {pages} {943} (\bibinfo {year} {2009})}\BibitemShut {NoStop}%
\bibitem [{\citenamefont {Kim}\ \emph {et~al.}(2016)\citenamefont {Kim},
  \citenamefont {Nomura}, \citenamefont {Ferrero}, \citenamefont {Seth},
  \citenamefont {Parcollet},\ and\ \citenamefont
  {Georges}}]{PhysRevB.94.155152}%
  \BibitemOpen
  \bibfield  {author} {\bibinfo {author} {\bibfnamefont {M.}~\bibnamefont
  {Kim}}, \bibinfo {author} {\bibfnamefont {Y.}~\bibnamefont {Nomura}},
  \bibinfo {author} {\bibfnamefont {M.}~\bibnamefont {Ferrero}}, \bibinfo
  {author} {\bibfnamefont {P.}~\bibnamefont {Seth}}, \bibinfo {author}
  {\bibfnamefont {O.}~\bibnamefont {Parcollet}}, \ and\ \bibinfo {author}
  {\bibfnamefont {A.}~\bibnamefont {Georges}},\ }\href {\doibase
  10.1103/PhysRevB.94.155152} {\bibfield  {journal} {\bibinfo  {journal} {Phys.
  Rev. B}\ }\textbf {\bibinfo {volume} {94}},\ \bibinfo {pages} {155152}
  (\bibinfo {year} {2016})}\BibitemShut {NoStop}%
\bibitem [{\citenamefont {Hoshino}\ and\ \citenamefont
  {Werner}(2017{\natexlab{a}})}]{PhysRevLett.118.177002}%
  \BibitemOpen
  \bibfield  {author} {\bibinfo {author} {\bibfnamefont {S.}~\bibnamefont
  {Hoshino}}\ and\ \bibinfo {author} {\bibfnamefont {P.}~\bibnamefont
  {Werner}},\ }\href {\doibase 10.1103/PhysRevLett.118.177002} {\bibfield
  {journal} {\bibinfo  {journal} {Phys. Rev. Lett.}\ }\textbf {\bibinfo
  {volume} {118}},\ \bibinfo {pages} {177002} (\bibinfo {year}
  {2017}{\natexlab{a}})}\BibitemShut {NoStop}%
\bibitem [{\citenamefont {Werner}\ \emph {et~al.}(2008)\citenamefont {Werner},
  \citenamefont {Gull}, \citenamefont {Troyer},\ and\ \citenamefont
  {Millis}}]{prl2008_Werner_spinfreezing}%
  \BibitemOpen
  \bibfield  {author} {\bibinfo {author} {\bibfnamefont {P.}~\bibnamefont
  {Werner}}, \bibinfo {author} {\bibfnamefont {E.}~\bibnamefont {Gull}},
  \bibinfo {author} {\bibfnamefont {M.}~\bibnamefont {Troyer}}, \ and\ \bibinfo
  {author} {\bibfnamefont {A.~J.}\ \bibnamefont {Millis}},\ }\href {\doibase
  10.1103/PhysRevLett.101.166405} {\bibfield  {journal} {\bibinfo  {journal}
  {Phys. Rev. Lett.}\ }\textbf {\bibinfo {volume} {101}},\ \bibinfo {pages}
  {166405} (\bibinfo {year} {2008})}\BibitemShut {NoStop}%
\bibitem [{\citenamefont {Chan}\ \emph {et~al.}(2009)\citenamefont {Chan},
  \citenamefont {Werner},\ and\ \citenamefont {Millis}}]{Chan2009}%
  \BibitemOpen
  \bibfield  {author} {\bibinfo {author} {\bibfnamefont {C.-K.}\ \bibnamefont
  {Chan}}, \bibinfo {author} {\bibfnamefont {P.}~\bibnamefont {Werner}}, \ and\
  \bibinfo {author} {\bibfnamefont {A.~J.}\ \bibnamefont {Millis}},\ }\href
  {\doibase 10.1103/PhysRevB.80.235114} {\bibfield  {journal} {\bibinfo
  {journal} {Phys. Rev. B}\ }\textbf {\bibinfo {volume} {80}},\ \bibinfo
  {pages} {235114} (\bibinfo {year} {2009})}\BibitemShut {NoStop}%
\bibitem [{\citenamefont {Werner}\ \emph {et~al.}(2009)\citenamefont {Werner},
  \citenamefont {Gull},\ and\ \citenamefont {Millis}}]{Werner2009}%
  \BibitemOpen
  \bibfield  {author} {\bibinfo {author} {\bibfnamefont {P.}~\bibnamefont
  {Werner}}, \bibinfo {author} {\bibfnamefont {E.}~\bibnamefont {Gull}}, \ and\
  \bibinfo {author} {\bibfnamefont {A.~J.}\ \bibnamefont {Millis}},\ }\href
  {\doibase 10.1103/PhysRevB.79.115119} {\bibfield  {journal} {\bibinfo
  {journal} {Phys. Rev. B}\ }\textbf {\bibinfo {volume} {79}},\ \bibinfo
  {pages} {115119} (\bibinfo {year} {2009})}\BibitemShut {NoStop}%
\bibitem [{\citenamefont {Kita}\ \emph {et~al.}(2011)\citenamefont {Kita},
  \citenamefont {Ohashi},\ and\ \citenamefont {Kawakami}}]{Kita2011}%
  \BibitemOpen
  \bibfield  {author} {\bibinfo {author} {\bibfnamefont {T.}~\bibnamefont
  {Kita}}, \bibinfo {author} {\bibfnamefont {T.}~\bibnamefont {Ohashi}}, \ and\
  \bibinfo {author} {\bibfnamefont {N.}~\bibnamefont {Kawakami}},\ }\href
  {\doibase 10.1103/PhysRevB.84.195130} {\bibfield  {journal} {\bibinfo
  {journal} {Phys. Rev. B}\ }\textbf {\bibinfo {volume} {84}},\ \bibinfo
  {pages} {195130} (\bibinfo {year} {2011})}\BibitemShut {NoStop}%
\bibitem [{\citenamefont {de' Medici}\ \emph {et~al.}(2011)\citenamefont {de'
  Medici}, \citenamefont {Mravlje},\ and\ \citenamefont
  {Georges}}]{PhysRevLett.107.256401}%
  \BibitemOpen
  \bibfield  {author} {\bibinfo {author} {\bibfnamefont {L.}~\bibnamefont {de'
  Medici}}, \bibinfo {author} {\bibfnamefont {J.}~\bibnamefont {Mravlje}}, \
  and\ \bibinfo {author} {\bibfnamefont {A.}~\bibnamefont {Georges}},\ }\href
  {\doibase 10.1103/PhysRevLett.107.256401} {\bibfield  {journal} {\bibinfo
  {journal} {Phys. Rev. Lett.}\ }\textbf {\bibinfo {volume} {107}},\ \bibinfo
  {pages} {256401} (\bibinfo {year} {2011})}\BibitemShut {NoStop}%
\bibitem [{\citenamefont {Okada}\ and\ \citenamefont
  {Yosida}(1973)}]{Okada_sd_quench_1973}%
  \BibitemOpen
  \bibfield  {author} {\bibinfo {author} {\bibfnamefont {I.}~\bibnamefont
  {Okada}}\ and\ \bibinfo {author} {\bibfnamefont {K.}~\bibnamefont {Yosida}},\
  }\href {\doibase 10.1143/PTP.49.1483} {\bibfield  {journal} {\bibinfo
  {journal} {Progress of Theoretical Physics}\ }\textbf {\bibinfo {volume}
  {49}},\ \bibinfo {pages} {1483} (\bibinfo {year} {1973})},\ \Eprint
  {http://arxiv.org/abs/https://academic.oup.com/ptp/article-pdf/49/5/1483/5410043/49-5-1483.pdf}
  {https://academic.oup.com/ptp/article-pdf/49/5/1483/5410043/49-5-1483.pdf}
  \BibitemShut {NoStop}%
\bibitem [{\citenamefont {Jayaprakash}\ \emph {et~al.}(1981)\citenamefont
  {Jayaprakash}, \citenamefont {Krishna-murthy},\ and\ \citenamefont
  {Wilkins}}]{PhysRevLett.47.737}%
  \BibitemOpen
  \bibfield  {author} {\bibinfo {author} {\bibfnamefont {C.}~\bibnamefont
  {Jayaprakash}}, \bibinfo {author} {\bibfnamefont {H.~R.}\ \bibnamefont
  {Krishna-murthy}}, \ and\ \bibinfo {author} {\bibfnamefont {J.~W.}\
  \bibnamefont {Wilkins}},\ }\href {\doibase 10.1103/PhysRevLett.47.737}
  {\bibfield  {journal} {\bibinfo  {journal} {Phys. Rev. Lett.}\ }\textbf
  {\bibinfo {volume} {47}},\ \bibinfo {pages} {737} (\bibinfo {year}
  {1981})}\BibitemShut {NoStop}%
\bibitem [{\citenamefont {Jones}\ and\ \citenamefont
  {Varma}(1987)}]{PhysRevLett.58.843}%
  \BibitemOpen
  \bibfield  {author} {\bibinfo {author} {\bibfnamefont {B.~A.}\ \bibnamefont
  {Jones}}\ and\ \bibinfo {author} {\bibfnamefont {C.~M.}\ \bibnamefont
  {Varma}},\ }\href {\doibase 10.1103/PhysRevLett.58.843} {\bibfield  {journal}
  {\bibinfo  {journal} {Phys. Rev. Lett.}\ }\textbf {\bibinfo {volume} {58}},\
  \bibinfo {pages} {843} (\bibinfo {year} {1987})}\BibitemShut {NoStop}%
\bibitem [{\citenamefont {Kusunose}\ and\ \citenamefont
  {Miyake}(1997)}]{JPSJ.66.1180}%
  \BibitemOpen
  \bibfield  {author} {\bibinfo {author} {\bibfnamefont {H.}~\bibnamefont
  {Kusunose}}\ and\ \bibinfo {author} {\bibfnamefont {K.}~\bibnamefont
  {Miyake}},\ }\href {\doibase 10.1143/JPSJ.66.1180} {\bibfield  {journal}
  {\bibinfo  {journal} {Journal of the Physical Society of Japan}\ }\textbf
  {\bibinfo {volume} {66}},\ \bibinfo {pages} {1180} (\bibinfo {year}
  {1997})},\ \Eprint
  {http://arxiv.org/abs/https://journals.jps.jp/doi/pdf/10.1143/JPSJ.66.1180}
  {https://journals.jps.jp/doi/pdf/10.1143/JPSJ.66.1180} \BibitemShut {NoStop}%
\bibitem [{\citenamefont {Daybell}\ and\ \citenamefont
  {Steyert}(1968)}]{RevModPhys.40.380}%
  \BibitemOpen
  \bibfield  {author} {\bibinfo {author} {\bibfnamefont {M.~D.}\ \bibnamefont
  {Daybell}}\ and\ \bibinfo {author} {\bibfnamefont {W.~A.}\ \bibnamefont
  {Steyert}},\ }\href {\doibase 10.1103/RevModPhys.40.380} {\bibfield
  {journal} {\bibinfo  {journal} {Rev. Mod. Phys.}\ }\textbf {\bibinfo {volume}
  {40}},\ \bibinfo {pages} {380} (\bibinfo {year} {1968})}\BibitemShut
  {NoStop}%
\bibitem [{\citenamefont {Nevidomskyy}\ and\ \citenamefont
  {Coleman}(2009)}]{Nevidomskyy2009}%
  \BibitemOpen
  \bibfield  {author} {\bibinfo {author} {\bibfnamefont {A.~H.}\ \bibnamefont
  {Nevidomskyy}}\ and\ \bibinfo {author} {\bibfnamefont {P.}~\bibnamefont
  {Coleman}},\ }\href {\doibase 10.1103/PhysRevLett.103.147205} {\bibfield
  {journal} {\bibinfo  {journal} {Phys. Rev. Lett.}\ }\textbf {\bibinfo
  {volume} {103}},\ \bibinfo {pages} {147205} (\bibinfo {year}
  {2009})}\BibitemShut {NoStop}%
\bibitem [{\citenamefont {Hoshino}\ and\ \citenamefont
  {Werner}(2015)}]{Hoshino2015}%
  \BibitemOpen
  \bibfield  {author} {\bibinfo {author} {\bibfnamefont {S.}~\bibnamefont
  {Hoshino}}\ and\ \bibinfo {author} {\bibfnamefont {P.}~\bibnamefont
  {Werner}},\ }\href {\doibase 10.1103/PhysRevLett.115.247001} {\bibfield
  {journal} {\bibinfo  {journal} {Phys. Rev. Lett.}\ }\textbf {\bibinfo
  {volume} {115}},\ \bibinfo {pages} {247001} (\bibinfo {year}
  {2015})}\BibitemShut {NoStop}%
\bibitem [{\citenamefont {Tyler}\ \emph {et~al.}(1998)\citenamefont {Tyler},
  \citenamefont {Mackenzie}, \citenamefont {NishiZaki},\ and\ \citenamefont
  {Maeno}}]{PhysRevB.58.R10107}%
  \BibitemOpen
  \bibfield  {author} {\bibinfo {author} {\bibfnamefont {A.~W.}\ \bibnamefont
  {Tyler}}, \bibinfo {author} {\bibfnamefont {A.~P.}\ \bibnamefont
  {Mackenzie}}, \bibinfo {author} {\bibfnamefont {S.}~\bibnamefont
  {NishiZaki}}, \ and\ \bibinfo {author} {\bibfnamefont {Y.}~\bibnamefont
  {Maeno}},\ }\href {\doibase 10.1103/PhysRevB.58.R10107} {\bibfield  {journal}
  {\bibinfo  {journal} {Phys. Rev. B}\ }\textbf {\bibinfo {volume} {58}},\
  \bibinfo {pages} {R10107} (\bibinfo {year} {1998})}\BibitemShut {NoStop}%
\bibitem [{\citenamefont {Yin}\ \emph {et~al.}(2012)\citenamefont {Yin},
  \citenamefont {Haule},\ and\ \citenamefont {Kotliar}}]{PhysRevB.86.195141}%
  \BibitemOpen
  \bibfield  {author} {\bibinfo {author} {\bibfnamefont {Z.~P.}\ \bibnamefont
  {Yin}}, \bibinfo {author} {\bibfnamefont {K.}~\bibnamefont {Haule}}, \ and\
  \bibinfo {author} {\bibfnamefont {G.}~\bibnamefont {Kotliar}},\ }\href
  {\doibase 10.1103/PhysRevB.86.195141} {\bibfield  {journal} {\bibinfo
  {journal} {Phys. Rev. B}\ }\textbf {\bibinfo {volume} {86}},\ \bibinfo
  {pages} {195141} (\bibinfo {year} {2012})}\BibitemShut {NoStop}%
\bibitem [{\citenamefont {Dang}\ \emph {et~al.}(2015)\citenamefont {Dang},
  \citenamefont {Mravlje}, \citenamefont {Georges},\ and\ \citenamefont
  {Millis}}]{PhysRevB.91.195149}%
  \BibitemOpen
  \bibfield  {author} {\bibinfo {author} {\bibfnamefont {H.~T.}\ \bibnamefont
  {Dang}}, \bibinfo {author} {\bibfnamefont {J.}~\bibnamefont {Mravlje}},
  \bibinfo {author} {\bibfnamefont {A.}~\bibnamefont {Georges}}, \ and\
  \bibinfo {author} {\bibfnamefont {A.~J.}\ \bibnamefont {Millis}},\ }\href
  {\doibase 10.1103/PhysRevB.91.195149} {\bibfield  {journal} {\bibinfo
  {journal} {Phys. Rev. B}\ }\textbf {\bibinfo {volume} {91}},\ \bibinfo
  {pages} {195149} (\bibinfo {year} {2015})}\BibitemShut {NoStop}%
\bibitem [{\citenamefont {Schneider}\ \emph {et~al.}(2014)\citenamefont
  {Schneider}, \citenamefont {Geiger}, \citenamefont {Esser}, \citenamefont
  {Pracht}, \citenamefont {Stingl}, \citenamefont {Tokiwa}, \citenamefont
  {Moshnyaga}, \citenamefont {Sheikin}, \citenamefont {Mravlje}, \citenamefont
  {Scheffler},\ and\ \citenamefont {Gegenwart}}]{PhysRevLett.112.206403}%
  \BibitemOpen
  \bibfield  {author} {\bibinfo {author} {\bibfnamefont {M.}~\bibnamefont
  {Schneider}}, \bibinfo {author} {\bibfnamefont {D.}~\bibnamefont {Geiger}},
  \bibinfo {author} {\bibfnamefont {S.}~\bibnamefont {Esser}}, \bibinfo
  {author} {\bibfnamefont {U.~S.}\ \bibnamefont {Pracht}}, \bibinfo {author}
  {\bibfnamefont {C.}~\bibnamefont {Stingl}}, \bibinfo {author} {\bibfnamefont
  {Y.}~\bibnamefont {Tokiwa}}, \bibinfo {author} {\bibfnamefont
  {V.}~\bibnamefont {Moshnyaga}}, \bibinfo {author} {\bibfnamefont
  {I.}~\bibnamefont {Sheikin}}, \bibinfo {author} {\bibfnamefont
  {J.}~\bibnamefont {Mravlje}}, \bibinfo {author} {\bibfnamefont
  {M.}~\bibnamefont {Scheffler}}, \ and\ \bibinfo {author} {\bibfnamefont
  {P.}~\bibnamefont {Gegenwart}},\ }\href {\doibase
  10.1103/PhysRevLett.112.206403} {\bibfield  {journal} {\bibinfo  {journal}
  {Phys. Rev. Lett.}\ }\textbf {\bibinfo {volume} {112}},\ \bibinfo {pages}
  {206403} (\bibinfo {year} {2014})}\BibitemShut {NoStop}%
\bibitem [{\citenamefont {Deng}\ \emph {et~al.}(2016)\citenamefont {Deng},
  \citenamefont {Haule},\ and\ \citenamefont
  {Kotliar}}]{PhysRevLett.116.256401}%
  \BibitemOpen
  \bibfield  {author} {\bibinfo {author} {\bibfnamefont {X.}~\bibnamefont
  {Deng}}, \bibinfo {author} {\bibfnamefont {K.}~\bibnamefont {Haule}}, \ and\
  \bibinfo {author} {\bibfnamefont {G.}~\bibnamefont {Kotliar}},\ }\href
  {\doibase 10.1103/PhysRevLett.116.256401} {\bibfield  {journal} {\bibinfo
  {journal} {Phys. Rev. Lett.}\ }\textbf {\bibinfo {volume} {116}},\ \bibinfo
  {pages} {256401} (\bibinfo {year} {2016})}\BibitemShut {NoStop}%
\bibitem [{\citenamefont {Ishida}\ and\ \citenamefont
  {Liebsch}(2010)}]{PhysRevB.81.054513}%
  \BibitemOpen
  \bibfield  {author} {\bibinfo {author} {\bibfnamefont {H.}~\bibnamefont
  {Ishida}}\ and\ \bibinfo {author} {\bibfnamefont {A.}~\bibnamefont
  {Liebsch}},\ }\href {\doibase 10.1103/PhysRevB.81.054513} {\bibfield
  {journal} {\bibinfo  {journal} {Phys. Rev. B}\ }\textbf {\bibinfo {volume}
  {81}},\ \bibinfo {pages} {054513} (\bibinfo {year} {2010})}\BibitemShut
  {NoStop}%
\bibitem [{\citenamefont {Ong}\ and\ \citenamefont
  {Coleman}(2012)}]{PhysRevLett.108.107201}%
  \BibitemOpen
  \bibfield  {author} {\bibinfo {author} {\bibfnamefont {T.~T.}\ \bibnamefont
  {Ong}}\ and\ \bibinfo {author} {\bibfnamefont {P.}~\bibnamefont {Coleman}},\
  }\href {\doibase 10.1103/PhysRevLett.108.107201} {\bibfield  {journal}
  {\bibinfo  {journal} {Phys. Rev. Lett.}\ }\textbf {\bibinfo {volume} {108}},\
  \bibinfo {pages} {107201} (\bibinfo {year} {2012})}\BibitemShut {NoStop}%
\bibitem [{\citenamefont {Xu}\ \emph {et~al.}(2013)\citenamefont {Xu},
  \citenamefont {Richard}, \citenamefont {van Roekeghem}, \citenamefont
  {Zhang}, \citenamefont {Miao}, \citenamefont {Zhang}, \citenamefont {Qian},
  \citenamefont {Ferrero}, \citenamefont {Sefat}, \citenamefont {Biermann},\
  and\ \citenamefont {Ding}}]{PhysRevX.3.011006}%
  \BibitemOpen
  \bibfield  {author} {\bibinfo {author} {\bibfnamefont {N.}~\bibnamefont
  {Xu}}, \bibinfo {author} {\bibfnamefont {P.}~\bibnamefont {Richard}},
  \bibinfo {author} {\bibfnamefont {A.}~\bibnamefont {van Roekeghem}}, \bibinfo
  {author} {\bibfnamefont {P.}~\bibnamefont {Zhang}}, \bibinfo {author}
  {\bibfnamefont {H.}~\bibnamefont {Miao}}, \bibinfo {author} {\bibfnamefont
  {W.-L.}\ \bibnamefont {Zhang}}, \bibinfo {author} {\bibfnamefont
  {T.}~\bibnamefont {Qian}}, \bibinfo {author} {\bibfnamefont {M.}~\bibnamefont
  {Ferrero}}, \bibinfo {author} {\bibfnamefont {A.~S.}\ \bibnamefont {Sefat}},
  \bibinfo {author} {\bibfnamefont {S.}~\bibnamefont {Biermann}}, \ and\
  \bibinfo {author} {\bibfnamefont {H.}~\bibnamefont {Ding}},\ }\href {\doibase
  10.1103/PhysRevX.3.011006} {\bibfield  {journal} {\bibinfo  {journal} {Phys.
  Rev. X}\ }\textbf {\bibinfo {volume} {3}},\ \bibinfo {pages} {011006}
  (\bibinfo {year} {2013})}\BibitemShut {NoStop}%
\bibitem [{\citenamefont {Haule}\ and\ \citenamefont
  {Kotliar}(2009)}]{Haule_2009}%
  \BibitemOpen
  \bibfield  {author} {\bibinfo {author} {\bibfnamefont {K.}~\bibnamefont
  {Haule}}\ and\ \bibinfo {author} {\bibfnamefont {G.}~\bibnamefont
  {Kotliar}},\ }\href {\doibase 10.1088/1367-2630/11/2/025021} {\bibfield
  {journal} {\bibinfo  {journal} {New Journal of Physics}\ }\textbf {\bibinfo
  {volume} {11}},\ \bibinfo {pages} {025021} (\bibinfo {year}
  {2009})}\BibitemShut {NoStop}%
\bibitem [{\citenamefont {Werner}\ \emph {et~al.}(2012)\citenamefont {Werner},
  \citenamefont {Casula}, \citenamefont {Miyake}, \citenamefont {Aryasetiawan},
  \citenamefont {Millis},\ and\ \citenamefont {Biermann}}]{Werner_NatPhys2012}%
  \BibitemOpen
  \bibfield  {author} {\bibinfo {author} {\bibfnamefont {P.}~\bibnamefont
  {Werner}}, \bibinfo {author} {\bibfnamefont {M.}~\bibnamefont {Casula}},
  \bibinfo {author} {\bibfnamefont {T.}~\bibnamefont {Miyake}}, \bibinfo
  {author} {\bibfnamefont {F.}~\bibnamefont {Aryasetiawan}}, \bibinfo {author}
  {\bibfnamefont {A.~J.}\ \bibnamefont {Millis}}, \ and\ \bibinfo {author}
  {\bibfnamefont {S.}~\bibnamefont {Biermann}},\ }\href {\doibase
  10.1038/nphys2250} {\bibfield  {journal} {\bibinfo  {journal} {Nature
  Physics}\ }\textbf {\bibinfo {volume} {8}},\ \bibinfo {pages} {331} (\bibinfo
  {year} {2012})}\BibitemShut {NoStop}%
\bibitem [{\citenamefont {Werner}\ \emph {et~al.}(2016)\citenamefont {Werner},
  \citenamefont {Hoshino},\ and\ \citenamefont
  {Shinaoka}}]{PhysRevB.94.245134}%
  \BibitemOpen
  \bibfield  {author} {\bibinfo {author} {\bibfnamefont {P.}~\bibnamefont
  {Werner}}, \bibinfo {author} {\bibfnamefont {S.}~\bibnamefont {Hoshino}}, \
  and\ \bibinfo {author} {\bibfnamefont {H.}~\bibnamefont {Shinaoka}},\ }\href
  {\doibase 10.1103/PhysRevB.94.245134} {\bibfield  {journal} {\bibinfo
  {journal} {Phys. Rev. B}\ }\textbf {\bibinfo {volume} {94}},\ \bibinfo
  {pages} {245134} (\bibinfo {year} {2016})}\BibitemShut {NoStop}%
\bibitem [{\citenamefont {Lenihan}\ \emph {et~al.}(2020)\citenamefont
  {Lenihan}, \citenamefont {Kim}, \citenamefont {Šimkovic IV.},\ and\
  \citenamefont {Kozik}}]{lenihan2020entropy}%
  \BibitemOpen
  \bibfield  {author} {\bibinfo {author} {\bibfnamefont {C.}~\bibnamefont
  {Lenihan}}, \bibinfo {author} {\bibfnamefont {A.~J.}\ \bibnamefont {Kim}},
  \bibinfo {author} {\bibfnamefont {F.}~\bibnamefont {Šimkovic IV.}}, \ and\
  \bibinfo {author} {\bibfnamefont {E.}~\bibnamefont {Kozik}},\ }\href@noop {}
  {\enquote {\bibinfo {title} {Entropy in the non-fermi-liquid regime of the
  doped $2d$ hubbard model},}\ } (\bibinfo {year} {2020}),\ \Eprint
  {http://arxiv.org/abs/2001.09948} {arXiv:2001.09948 [cond-mat.str-el]}
  \BibitemShut {NoStop}%
\bibitem [{\citenamefont {Ishigaki}\ \emph {et~al.}(2018)\citenamefont
  {Ishigaki}, \citenamefont {Nasu}, \citenamefont {Koga}, \citenamefont
  {Hoshino},\ and\ \citenamefont {Werner}}]{PhysRevB.98.235120}%
  \BibitemOpen
  \bibfield  {author} {\bibinfo {author} {\bibfnamefont {K.}~\bibnamefont
  {Ishigaki}}, \bibinfo {author} {\bibfnamefont {J.}~\bibnamefont {Nasu}},
  \bibinfo {author} {\bibfnamefont {A.}~\bibnamefont {Koga}}, \bibinfo {author}
  {\bibfnamefont {S.}~\bibnamefont {Hoshino}}, \ and\ \bibinfo {author}
  {\bibfnamefont {P.}~\bibnamefont {Werner}},\ }\href {\doibase
  10.1103/PhysRevB.98.235120} {\bibfield  {journal} {\bibinfo  {journal} {Phys.
  Rev. B}\ }\textbf {\bibinfo {volume} {98}},\ \bibinfo {pages} {235120}
  (\bibinfo {year} {2018})}\BibitemShut {NoStop}%
\bibitem [{\citenamefont {Stadler}\ \emph {et~al.}(2019)\citenamefont
  {Stadler}, \citenamefont {Kotliar}, \citenamefont {Weichselbaum},\ and\
  \citenamefont {von Delft}}]{STADLER2019365}%
  \BibitemOpen
  \bibfield  {author} {\bibinfo {author} {\bibfnamefont {K.}~\bibnamefont
  {Stadler}}, \bibinfo {author} {\bibfnamefont {G.}~\bibnamefont {Kotliar}},
  \bibinfo {author} {\bibfnamefont {A.}~\bibnamefont {Weichselbaum}}, \ and\
  \bibinfo {author} {\bibfnamefont {J.}~\bibnamefont {von Delft}},\ }\href
  {\doibase https://doi.org/10.1016/j.aop.2018.10.017} {\bibfield  {journal}
  {\bibinfo  {journal} {Annals of Physics}\ }\textbf {\bibinfo {volume}
  {405}},\ \bibinfo {pages} {365 } (\bibinfo {year} {2019})}\BibitemShut
  {NoStop}%
\bibitem [{\citenamefont {Stadler}\ \emph {et~al.}(2015)\citenamefont
  {Stadler}, \citenamefont {Yin}, \citenamefont {von Delft}, \citenamefont
  {Kotliar},\ and\ \citenamefont {Weichselbaum}}]{PhysRevLett.115.136401}%
  \BibitemOpen
  \bibfield  {author} {\bibinfo {author} {\bibfnamefont {K.~M.}\ \bibnamefont
  {Stadler}}, \bibinfo {author} {\bibfnamefont {Z.~P.}\ \bibnamefont {Yin}},
  \bibinfo {author} {\bibfnamefont {J.}~\bibnamefont {von Delft}}, \bibinfo
  {author} {\bibfnamefont {G.}~\bibnamefont {Kotliar}}, \ and\ \bibinfo
  {author} {\bibfnamefont {A.}~\bibnamefont {Weichselbaum}},\ }\href {\doibase
  10.1103/PhysRevLett.115.136401} {\bibfield  {journal} {\bibinfo  {journal}
  {Phys. Rev. Lett.}\ }\textbf {\bibinfo {volume} {115}},\ \bibinfo {pages}
  {136401} (\bibinfo {year} {2015})}\BibitemShut {NoStop}%
\bibitem [{\citenamefont {Deng}\ \emph {et~al.}(2019)\citenamefont {Deng},
  \citenamefont {Stadler}, \citenamefont {Haule}, \citenamefont {Weichselbaum},
  \citenamefont {von Delft},\ and\ \citenamefont
  {Kotliar}}]{XYDeng_NatComm_2019}%
  \BibitemOpen
  \bibfield  {author} {\bibinfo {author} {\bibfnamefont {X.}~\bibnamefont
  {Deng}}, \bibinfo {author} {\bibfnamefont {K.~M.}\ \bibnamefont {Stadler}},
  \bibinfo {author} {\bibfnamefont {K.}~\bibnamefont {Haule}}, \bibinfo
  {author} {\bibfnamefont {A.}~\bibnamefont {Weichselbaum}}, \bibinfo {author}
  {\bibfnamefont {J.}~\bibnamefont {von Delft}}, \ and\ \bibinfo {author}
  {\bibfnamefont {G.}~\bibnamefont {Kotliar}},\ }\href {\doibase
  10.1038/s41467-019-10257-2} {\bibfield  {journal} {\bibinfo  {journal} {Nat.
  Commun.}\ }\textbf {\bibinfo {volume} {10}},\ \bibinfo {pages} {2721}
  (\bibinfo {year} {2019})}\BibitemShut {NoStop}%
\bibitem [{\citenamefont {Steiner}\ \emph {et~al.}(2016)\citenamefont
  {Steiner}, \citenamefont {Hoshino}, \citenamefont {Nomura},\ and\
  \citenamefont {Werner}}]{Steiner2016}%
  \BibitemOpen
  \bibfield  {author} {\bibinfo {author} {\bibfnamefont {K.}~\bibnamefont
  {Steiner}}, \bibinfo {author} {\bibfnamefont {S.}~\bibnamefont {Hoshino}},
  \bibinfo {author} {\bibfnamefont {Y.}~\bibnamefont {Nomura}}, \ and\ \bibinfo
  {author} {\bibfnamefont {P.}~\bibnamefont {Werner}},\ }\href {\doibase
  10.1103/PhysRevB.94.075107} {\bibfield  {journal} {\bibinfo  {journal} {Phys.
  Rev. B}\ }\textbf {\bibinfo {volume} {94}},\ \bibinfo {pages} {075107}
  (\bibinfo {year} {2016})}\BibitemShut {NoStop}%
\bibitem [{\citenamefont {Hoshino}\ and\ \citenamefont
  {Werner}(2016)}]{Hoshino2016}%
  \BibitemOpen
  \bibfield  {author} {\bibinfo {author} {\bibfnamefont {S.}~\bibnamefont
  {Hoshino}}\ and\ \bibinfo {author} {\bibfnamefont {P.}~\bibnamefont
  {Werner}},\ }\href {\doibase 10.1103/PhysRevB.93.155161} {\bibfield
  {journal} {\bibinfo  {journal} {Phys. Rev. B}\ }\textbf {\bibinfo {volume}
  {93}},\ \bibinfo {pages} {155161} (\bibinfo {year} {2016})}\BibitemShut
  {NoStop}%
\bibitem [{\citenamefont {{Horvat}}\ \emph {et~al.}(2019)\citenamefont
  {{Horvat}}, \citenamefont {{Zitko}},\ and\ \citenamefont
  {{Mravlje}}}]{2019arXiv190707100H}%
  \BibitemOpen
  \bibfield  {author} {\bibinfo {author} {\bibfnamefont {A.}~\bibnamefont
  {{Horvat}}}, \bibinfo {author} {\bibfnamefont {R.}~\bibnamefont {{Zitko}}}, \
  and\ \bibinfo {author} {\bibfnamefont {J.}~\bibnamefont {{Mravlje}}},\
  }\href@noop {} {\bibfield  {journal} {\bibinfo  {journal} {arXiv e-prints}\
  ,\ \bibinfo {eid} {arXiv:1907.07100}} (\bibinfo {year} {2019})},\ \Eprint
  {http://arxiv.org/abs/1907.07100} {arXiv:1907.07100 [cond-mat.str-el]}
  \BibitemShut {NoStop}%
\bibitem [{\citenamefont {{Wang}}\ \emph {et~al.}(2019)\citenamefont {{Wang}},
  \citenamefont {{Walter}}, \citenamefont {{Lee}}, \citenamefont {{Stadler}},
  \citenamefont {{von Delft}}, \citenamefont {{Weichselbaum}},\ and\
  \citenamefont {{Kotliar}}}]{2019arXiv191013643W}%
  \BibitemOpen
  \bibfield  {author} {\bibinfo {author} {\bibfnamefont {Y.}~\bibnamefont
  {{Wang}}}, \bibinfo {author} {\bibfnamefont {E.}~\bibnamefont {{Walter}}},
  \bibinfo {author} {\bibfnamefont {S.~S.~B.}\ \bibnamefont {{Lee}}}, \bibinfo
  {author} {\bibfnamefont {K.~M.}\ \bibnamefont {{Stadler}}}, \bibinfo {author}
  {\bibfnamefont {J.}~\bibnamefont {{von Delft}}}, \bibinfo {author}
  {\bibfnamefont {A.}~\bibnamefont {{Weichselbaum}}}, \ and\ \bibinfo {author}
  {\bibfnamefont {G.}~\bibnamefont {{Kotliar}}},\ }\href@noop {} {\bibfield
  {journal} {\bibinfo  {journal} {arXiv e-prints}\ ,\ \bibinfo {eid}
  {arXiv:1910.13643}} (\bibinfo {year} {2019})},\ \Eprint
  {http://arxiv.org/abs/1910.13643} {arXiv:1910.13643 [cond-mat.str-el]}
  \BibitemShut {NoStop}%
\bibitem [{\citenamefont {Werner}\ \emph {et~al.}(2006)\citenamefont {Werner},
  \citenamefont {Comanac}, \citenamefont {de' Medici}, \citenamefont {Troyer},\
  and\ \citenamefont {Millis}}]{ctqmc_prl2006}%
  \BibitemOpen
  \bibfield  {author} {\bibinfo {author} {\bibfnamefont {P.}~\bibnamefont
  {Werner}}, \bibinfo {author} {\bibfnamefont {A.}~\bibnamefont {Comanac}},
  \bibinfo {author} {\bibfnamefont {L.}~\bibnamefont {de' Medici}}, \bibinfo
  {author} {\bibfnamefont {M.}~\bibnamefont {Troyer}}, \ and\ \bibinfo {author}
  {\bibfnamefont {A.~J.}\ \bibnamefont {Millis}},\ }\href {\doibase
  10.1103/PhysRevLett.97.076405} {\bibfield  {journal} {\bibinfo  {journal}
  {Phys. Rev. Lett.}\ }\textbf {\bibinfo {volume} {97}},\ \bibinfo {pages}
  {076405} (\bibinfo {year} {2006})}\BibitemShut {NoStop}%
\bibitem [{\citenamefont {Gull}\ \emph {et~al.}(2011)\citenamefont {Gull},
  \citenamefont {Millis}, \citenamefont {Lichtenstein}, \citenamefont
  {Rubtsov}, \citenamefont {Troyer},\ and\ \citenamefont {Werner}}]{rmp_ctqmc}%
  \BibitemOpen
  \bibfield  {author} {\bibinfo {author} {\bibfnamefont {E.}~\bibnamefont
  {Gull}}, \bibinfo {author} {\bibfnamefont {A.~J.}\ \bibnamefont {Millis}},
  \bibinfo {author} {\bibfnamefont {A.~I.}\ \bibnamefont {Lichtenstein}},
  \bibinfo {author} {\bibfnamefont {A.~N.}\ \bibnamefont {Rubtsov}}, \bibinfo
  {author} {\bibfnamefont {M.}~\bibnamefont {Troyer}}, \ and\ \bibinfo {author}
  {\bibfnamefont {P.}~\bibnamefont {Werner}},\ }\href {\doibase
  10.1103/RevModPhys.83.349} {\bibfield  {journal} {\bibinfo  {journal} {Rev.
  Mod. Phys.}\ }\textbf {\bibinfo {volume} {83}},\ \bibinfo {pages} {349}
  (\bibinfo {year} {2011})}\BibitemShut {NoStop}%
\bibitem [{\citenamefont {Georges}\ \emph {et~al.}(1996)\citenamefont
  {Georges}, \citenamefont {Kotliar}, \citenamefont {Krauth},\ and\
  \citenamefont {Rozenberg}}]{rmp_68_13_dmft_1996}%
  \BibitemOpen
  \bibfield  {author} {\bibinfo {author} {\bibfnamefont {A.}~\bibnamefont
  {Georges}}, \bibinfo {author} {\bibfnamefont {G.}~\bibnamefont {Kotliar}},
  \bibinfo {author} {\bibfnamefont {W.}~\bibnamefont {Krauth}}, \ and\ \bibinfo
  {author} {\bibfnamefont {M.~J.}\ \bibnamefont {Rozenberg}},\ }\href {\doibase
  10.1103/RevModPhys.68.13} {\bibfield  {journal} {\bibinfo  {journal} {Rev.
  Mod. Phys.}\ }\textbf {\bibinfo {volume} {68}},\ \bibinfo {pages} {13}
  (\bibinfo {year} {1996})}\BibitemShut {NoStop}%
\bibitem [{\citenamefont {Ishigaki}\ \emph {et~al.}(2019)\citenamefont
  {Ishigaki}, \citenamefont {Nasu}, \citenamefont {Koga}, \citenamefont
  {Hoshino},\ and\ \citenamefont {Werner}}]{Ishigaki2019}%
  \BibitemOpen
  \bibfield  {author} {\bibinfo {author} {\bibfnamefont {K.}~\bibnamefont
  {Ishigaki}}, \bibinfo {author} {\bibfnamefont {J.}~\bibnamefont {Nasu}},
  \bibinfo {author} {\bibfnamefont {A.}~\bibnamefont {Koga}}, \bibinfo {author}
  {\bibfnamefont {S.}~\bibnamefont {Hoshino}}, \ and\ \bibinfo {author}
  {\bibfnamefont {P.}~\bibnamefont {Werner}},\ }\href {\doibase
  10.1103/PhysRevB.99.085131} {\bibfield  {journal} {\bibinfo  {journal} {Phys.
  Rev. B}\ }\textbf {\bibinfo {volume} {99}},\ \bibinfo {pages} {085131}
  (\bibinfo {year} {2019})}\BibitemShut {NoStop}%
\bibitem [{\citenamefont {Hoshino}\ and\ \citenamefont
  {Werner}(2017{\natexlab{b}})}]{Hoshino2017}%
  \BibitemOpen
  \bibfield  {author} {\bibinfo {author} {\bibfnamefont {S.}~\bibnamefont
  {Hoshino}}\ and\ \bibinfo {author} {\bibfnamefont {P.}~\bibnamefont
  {Werner}},\ }\href {\doibase 10.1103/PhysRevLett.118.177002} {\bibfield
  {journal} {\bibinfo  {journal} {Phys. Rev. Lett.}\ }\textbf {\bibinfo
  {volume} {118}},\ \bibinfo {pages} {177002} (\bibinfo {year}
  {2017}{\natexlab{b}})}\BibitemShut {NoStop}%
\bibitem [{\citenamefont {Haule}(2007)}]{prb2007_Haule_qmc}%
  \BibitemOpen
  \bibfield  {author} {\bibinfo {author} {\bibfnamefont {K.}~\bibnamefont
  {Haule}},\ }\href {\doibase 10.1103/PhysRevB.75.155113} {\bibfield  {journal}
  {\bibinfo  {journal} {Phys. Rev. B}\ }\textbf {\bibinfo {volume} {75}},\
  \bibinfo {pages} {155113} (\bibinfo {year} {2007})}\BibitemShut {NoStop}%
\bibitem [{\citenamefont {Duffy}\ and\ \citenamefont
  {Moreo}(1997)}]{PhysRevB.55.12918}%
  \BibitemOpen
  \bibfield  {author} {\bibinfo {author} {\bibfnamefont {D.}~\bibnamefont
  {Duffy}}\ and\ \bibinfo {author} {\bibfnamefont {A.}~\bibnamefont {Moreo}},\
  }\href {\doibase 10.1103/PhysRevB.55.12918} {\bibfield  {journal} {\bibinfo
  {journal} {Phys. Rev. B}\ }\textbf {\bibinfo {volume} {55}},\ \bibinfo
  {pages} {12918} (\bibinfo {year} {1997})}\BibitemShut {NoStop}%
\bibitem [{\citenamefont {Paiva}\ \emph {et~al.}(2001)\citenamefont {Paiva},
  \citenamefont {Huscroft}, \citenamefont {McMahan},\ and\ \citenamefont
  {Scalettar}}]{PAIVA2001224}%
  \BibitemOpen
  \bibfield  {author} {\bibinfo {author} {\bibfnamefont {T.}~\bibnamefont
  {Paiva}}, \bibinfo {author} {\bibfnamefont {C.}~\bibnamefont {Huscroft}},
  \bibinfo {author} {\bibfnamefont {A.}~\bibnamefont {McMahan}}, \ and\
  \bibinfo {author} {\bibfnamefont {R.}~\bibnamefont {Scalettar}},\ }\href
  {\doibase https://doi.org/10.1016/S0304-8853(00)00911-2} {\bibfield
  {journal} {\bibinfo  {journal} {Journal of Magnetism and Magnetic Materials}\
  }\textbf {\bibinfo {volume} {226-230}},\ \bibinfo {pages} {224 } (\bibinfo
  {year} {2001})},\ \bibinfo {note} {proceedings of the International
  Conference on Magnetism (ICM 2000)}\BibitemShut {NoStop}%
\bibitem [{\citenamefont {Huscroft}\ \emph {et~al.}(1999)\citenamefont
  {Huscroft}, \citenamefont {McMahan},\ and\ \citenamefont
  {Scalettar}}]{PhysRevLett.82.2342}%
  \BibitemOpen
  \bibfield  {author} {\bibinfo {author} {\bibfnamefont {C.}~\bibnamefont
  {Huscroft}}, \bibinfo {author} {\bibfnamefont {A.~K.}\ \bibnamefont
  {McMahan}}, \ and\ \bibinfo {author} {\bibfnamefont {R.~T.}\ \bibnamefont
  {Scalettar}},\ }\href {\doibase 10.1103/PhysRevLett.82.2342} {\bibfield
  {journal} {\bibinfo  {journal} {Phys. Rev. Lett.}\ }\textbf {\bibinfo
  {volume} {82}},\ \bibinfo {pages} {2342} (\bibinfo {year}
  {1999})}\BibitemShut {NoStop}%
\bibitem [{\citenamefont {Huang}\ \emph {et~al.}(2015)\citenamefont {Huang},
  \citenamefont {Wang}, \citenamefont {Meng}, \citenamefont {Du}, \citenamefont
  {Werner},\ and\ \citenamefont {Dai}}]{HUANG2015140}%
  \BibitemOpen
  \bibfield  {author} {\bibinfo {author} {\bibfnamefont {L.}~\bibnamefont
  {Huang}}, \bibinfo {author} {\bibfnamefont {Y.}~\bibnamefont {Wang}},
  \bibinfo {author} {\bibfnamefont {Z.~Y.}\ \bibnamefont {Meng}}, \bibinfo
  {author} {\bibfnamefont {L.}~\bibnamefont {Du}}, \bibinfo {author}
  {\bibfnamefont {P.}~\bibnamefont {Werner}}, \ and\ \bibinfo {author}
  {\bibfnamefont {X.}~\bibnamefont {Dai}},\ }\href {\doibase
  https://doi.org/10.1016/j.cpc.2015.04.020} {\bibfield  {journal} {\bibinfo
  {journal} {Computer Physics Communications}\ }\textbf {\bibinfo {volume}
  {195}},\ \bibinfo {pages} {140 } (\bibinfo {year} {2015})}\BibitemShut
  {NoStop}%
\bibitem [{\citenamefont {Huang}(2017)}]{iqist}%
  \BibitemOpen
  \bibfield  {author} {\bibinfo {author} {\bibfnamefont {L.}~\bibnamefont
  {Huang}},\ }\href {\doibase https://doi.org/10.1016/j.cpc.2017.08.026}
  {\bibfield  {journal} {\bibinfo  {journal} {Computer Physics Communications}\
  }\textbf {\bibinfo {volume} {221}},\ \bibinfo {pages} {423} (\bibinfo {year}
  {2017})}\BibitemShut {NoStop}%
\bibitem [{\citenamefont {Poteryaev}\ \emph {et~al.}(2007)\citenamefont
  {Poteryaev}, \citenamefont {Tomczak}, \citenamefont {Biermann}, \citenamefont
  {Georges}, \citenamefont {Lichtenstein}, \citenamefont {Rubtsov},
  \citenamefont {Saha-Dasgupta},\ and\ \citenamefont
  {Andersen}}]{PhysRevB.76.085127}%
  \BibitemOpen
  \bibfield  {author} {\bibinfo {author} {\bibfnamefont {A.~I.}\ \bibnamefont
  {Poteryaev}}, \bibinfo {author} {\bibfnamefont {J.~M.}\ \bibnamefont
  {Tomczak}}, \bibinfo {author} {\bibfnamefont {S.}~\bibnamefont {Biermann}},
  \bibinfo {author} {\bibfnamefont {A.}~\bibnamefont {Georges}}, \bibinfo
  {author} {\bibfnamefont {A.~I.}\ \bibnamefont {Lichtenstein}}, \bibinfo
  {author} {\bibfnamefont {A.~N.}\ \bibnamefont {Rubtsov}}, \bibinfo {author}
  {\bibfnamefont {T.}~\bibnamefont {Saha-Dasgupta}}, \ and\ \bibinfo {author}
  {\bibfnamefont {O.~K.}\ \bibnamefont {Andersen}},\ }\href {\doibase
  10.1103/PhysRevB.76.085127} {\bibfield  {journal} {\bibinfo  {journal} {Phys.
  Rev. B}\ }\textbf {\bibinfo {volume} {76}},\ \bibinfo {pages} {085127}
  (\bibinfo {year} {2007})}\BibitemShut {NoStop}%
\bibitem [{\citenamefont {Kowalski}\ \emph {et~al.}(2019)\citenamefont
  {Kowalski}, \citenamefont {Hausoel}, \citenamefont {Wallerberger},
  \citenamefont {Gunacker},\ and\ \citenamefont
  {Sangiovanni}}]{state_sampling_prb2019}%
  \BibitemOpen
  \bibfield  {author} {\bibinfo {author} {\bibfnamefont {A.}~\bibnamefont
  {Kowalski}}, \bibinfo {author} {\bibfnamefont {A.}~\bibnamefont {Hausoel}},
  \bibinfo {author} {\bibfnamefont {M.}~\bibnamefont {Wallerberger}}, \bibinfo
  {author} {\bibfnamefont {P.}~\bibnamefont {Gunacker}}, \ and\ \bibinfo
  {author} {\bibfnamefont {G.}~\bibnamefont {Sangiovanni}},\ }\href {\doibase
  10.1103/PhysRevB.99.155112} {\bibfield  {journal} {\bibinfo  {journal} {Phys.
  Rev. B}\ }\textbf {\bibinfo {volume} {99}},\ \bibinfo {pages} {155112}
  (\bibinfo {year} {2019})}\BibitemShut {NoStop}%
\bibitem [{\citenamefont {\ifmmode~\check{Z}\else
  \v{Z}\fi{}itko}(2009)}]{Broyden_mixing_Rok}%
  \BibitemOpen
  \bibfield  {author} {\bibinfo {author} {\bibfnamefont {R.}~\bibnamefont
  {\ifmmode~\check{Z}\else \v{Z}\fi{}itko}},\ }\href {\doibase
  10.1103/PhysRevB.80.125125} {\bibfield  {journal} {\bibinfo  {journal} {Phys.
  Rev. B}\ }\textbf {\bibinfo {volume} {80}},\ \bibinfo {pages} {125125}
  (\bibinfo {year} {2009})}\BibitemShut {NoStop}%
\bibitem [{\citenamefont {Werner}\ and\ \citenamefont
  {Millis}(2007{\natexlab{a}})}]{Werner2007doped}%
  \BibitemOpen
  \bibfield  {author} {\bibinfo {author} {\bibfnamefont {P.}~\bibnamefont
  {Werner}}\ and\ \bibinfo {author} {\bibfnamefont {A.~J.}\ \bibnamefont
  {Millis}},\ }\href {\doibase 10.1103/PhysRevB.75.085108} {\bibfield
  {journal} {\bibinfo  {journal} {Phys. Rev. B}\ }\textbf {\bibinfo {volume}
  {75}},\ \bibinfo {pages} {085108} (\bibinfo {year}
  {2007}{\natexlab{a}})}\BibitemShut {NoStop}%
\bibitem [{\citenamefont {Huang}\ \emph {et~al.}(2016)\citenamefont {Huang},
  \citenamefont {Wang}, \citenamefont {Wang},\ and\ \citenamefont
  {Werner}}]{Huang2016}%
  \BibitemOpen
  \bibfield  {author} {\bibinfo {author} {\bibfnamefont {L.}~\bibnamefont
  {Huang}}, \bibinfo {author} {\bibfnamefont {Y.}~\bibnamefont {Wang}},
  \bibinfo {author} {\bibfnamefont {L.}~\bibnamefont {Wang}}, \ and\ \bibinfo
  {author} {\bibfnamefont {P.}~\bibnamefont {Werner}},\ }\href {\doibase
  10.1103/PhysRevB.94.235110} {\bibfield  {journal} {\bibinfo  {journal} {Phys.
  Rev. B}\ }\textbf {\bibinfo {volume} {94}},\ \bibinfo {pages} {235110}
  (\bibinfo {year} {2016})}\BibitemShut {NoStop}%
\bibitem [{\citenamefont {Ellen}\ and\ \citenamefont
  {David}(1975)}]{PhysRevB.12.2260}%
  \BibitemOpen
  \bibfield  {author} {\bibinfo {author} {\bibfnamefont {J.~Y.}\ \bibnamefont
  {Ellen}}\ and\ \bibinfo {author} {\bibfnamefont {A.}~\bibnamefont {David}},\
  }\href {\doibase 10.1103/PhysRevB.12.2260} {\bibfield  {journal} {\bibinfo
  {journal} {Phys. Rev. B}\ }\textbf {\bibinfo {volume} {12}},\ \bibinfo
  {pages} {2260} (\bibinfo {year} {1975})}\BibitemShut {NoStop}%
\bibitem [{\citenamefont {Merker}\ \emph {et~al.}(2012)\citenamefont {Merker},
  \citenamefont {Weichselbaum},\ and\ \citenamefont
  {Costi}}]{PhysRevB.86.075153}%
  \BibitemOpen
  \bibfield  {author} {\bibinfo {author} {\bibfnamefont {L.}~\bibnamefont
  {Merker}}, \bibinfo {author} {\bibfnamefont {A.}~\bibnamefont
  {Weichselbaum}}, \ and\ \bibinfo {author} {\bibfnamefont {T.~A.}\
  \bibnamefont {Costi}},\ }\href {\doibase 10.1103/PhysRevB.86.075153}
  {\bibfield  {journal} {\bibinfo  {journal} {Phys. Rev. B}\ }\textbf {\bibinfo
  {volume} {86}},\ \bibinfo {pages} {075153} (\bibinfo {year}
  {2012})}\BibitemShut {NoStop}%
\bibitem [{\citenamefont {Hubbard}\ and\ \citenamefont
  {Flowers}(1964)}]{HubI_1964}%
  \BibitemOpen
  \bibfield  {author} {\bibinfo {author} {\bibfnamefont {J.}~\bibnamefont
  {Hubbard}}\ and\ \bibinfo {author} {\bibfnamefont {B.~H.}\ \bibnamefont
  {Flowers}},\ }\href {\doibase 10.1098/rspa.1964.0190} {\bibfield  {journal}
  {\bibinfo  {journal} {Proceedings of the Royal Society of London. Series A.
  Mathematical and Physical Sciences}\ }\textbf {\bibinfo {volume} {281}},\
  \bibinfo {pages} {401} (\bibinfo {year} {1964})}\BibitemShut {NoStop}%
\bibitem [{\citenamefont {Werner}\ and\ \citenamefont
  {Millis}(2007{\natexlab{b}})}]{prb2007_Werner_dopeMott}%
  \BibitemOpen
  \bibfield  {author} {\bibinfo {author} {\bibfnamefont {P.}~\bibnamefont
  {Werner}}\ and\ \bibinfo {author} {\bibfnamefont {A.~J.}\ \bibnamefont
  {Millis}},\ }\href {\doibase 10.1103/PhysRevB.75.085108} {\bibfield
  {journal} {\bibinfo  {journal} {Phys. Rev. B}\ }\textbf {\bibinfo {volume}
  {75}},\ \bibinfo {pages} {085108} (\bibinfo {year}
  {2007}{\natexlab{b}})}\BibitemShut {NoStop}%
\bibitem [{\citenamefont {Boehnke}\ \emph {et~al.}(2011)\citenamefont
  {Boehnke}, \citenamefont {Hafermann}, \citenamefont {Ferrero}, \citenamefont
  {Lechermann},\ and\ \citenamefont {Parcollet}}]{Lewin_prb_2011_polynomial}%
  \BibitemOpen
  \bibfield  {author} {\bibinfo {author} {\bibfnamefont {L.}~\bibnamefont
  {Boehnke}}, \bibinfo {author} {\bibfnamefont {H.}~\bibnamefont {Hafermann}},
  \bibinfo {author} {\bibfnamefont {M.}~\bibnamefont {Ferrero}}, \bibinfo
  {author} {\bibfnamefont {F.}~\bibnamefont {Lechermann}}, \ and\ \bibinfo
  {author} {\bibfnamefont {O.}~\bibnamefont {Parcollet}},\ }\href {\doibase
  10.1103/PhysRevB.84.075145} {\bibfield  {journal} {\bibinfo  {journal} {Phys.
  Rev. B}\ }\textbf {\bibinfo {volume} {84}},\ \bibinfo {pages} {075145}
  (\bibinfo {year} {2011})}\BibitemShut {NoStop}%
\bibitem [{\citenamefont {Gunacker}\ \emph {et~al.}(2015)\citenamefont
  {Gunacker}, \citenamefont {Wallerberger}, \citenamefont {Gull}, \citenamefont
  {Hausoel}, \citenamefont {Sangiovanni},\ and\ \citenamefont
  {Held}}]{ctqmc_worm_prb2015}%
  \BibitemOpen
  \bibfield  {author} {\bibinfo {author} {\bibfnamefont {P.}~\bibnamefont
  {Gunacker}}, \bibinfo {author} {\bibfnamefont {M.}~\bibnamefont
  {Wallerberger}}, \bibinfo {author} {\bibfnamefont {E.}~\bibnamefont {Gull}},
  \bibinfo {author} {\bibfnamefont {A.}~\bibnamefont {Hausoel}}, \bibinfo
  {author} {\bibfnamefont {G.}~\bibnamefont {Sangiovanni}}, \ and\ \bibinfo
  {author} {\bibfnamefont {K.}~\bibnamefont {Held}},\ }\href {\doibase
  10.1103/PhysRevB.92.155102} {\bibfield  {journal} {\bibinfo  {journal} {Phys.
  Rev. B}\ }\textbf {\bibinfo {volume} {92}},\ \bibinfo {pages} {155102}
  (\bibinfo {year} {2015})}\BibitemShut {NoStop}%
\end{thebibliography}%

\end{document}